\documentclass[sigconf,screen]{acmart}

\usepackage{algpseudocode}

\setcopyright{acmlicensed}
\acmDOI{10.1145/3663533.3664040}
\acmYear{2024}
\copyrightyear{2024}
\acmSubmissionID{fsews24promisemain-id15-p}
\acmISBN{979-8-4007-0675-2/24/07}
\acmConference[PROMISE '24]{Proceedings of the 20th International Conference on Predictive Models and Data Analytics in Software Engineering}{July 16, 2024}{Porto de Galinhas, Brazil}
\acmBooktitle{Proceedings of the 20th International Conference on Predictive Models and Data Analytics in Software Engineering (PROMISE '24), July 16, 2024, Porto de Galinhas, Brazil}
\usepackage{array}
\usepackage[caption=false,font=normalsize,labelfont=sf,textfont=sf]{subfig}
\usepackage{textcomp}
\usepackage{stfloats}
\usepackage{url}
\usepackage{verbatim}
\usepackage{graphicx}
\def\BibTeX{{\rm B\kern-.05em{\sc i\kern-.025em b}\kern-.08em
    T\kern-.1667em\lower.7ex\hbox{E}\kern-.125emX}}
\usepackage{balance}

\usepackage{amsthm}
\usepackage{mdframed}
\usepackage{centernot}
\usepackage{comment}

\newcommand{\Ff}{\mathcal{F}}

\newcommand{\Hh}{\mathcal{H}}
\newcommand{\Nn}{\mathcal{N}}

\newcommand{\Xx}{\mathcal{X}}
\newcommand{\Yy}{\mathcal{Y}}

\newcommand{\Dd}{\mathcal{D}}

\newcommand{\vmsays}[1]{{\color{cyan}{\bf VM:} #1}}

\newcommand\vd[2]{d_{i, p}}
\newcommand{\set}[1]{\left\{ #1 \right\}}

\newcommand{\R}{\mathbb R}

\newcommand{\Real}{\R}

\usepackage{tikz}

\definecolor{gold}{rgb}{0.99,0.78,0.07}
\usetikzlibrary{arrows,shapes,snakes,automata,backgrounds,positioning,decorations.pathmorphing,
	decorations.markings,calc}

\tikzstyle{dtreenode}=[draw=blue!10!gray,rounded rectangle, minimum size=5mm,fill=blue!10!white]
\tikzstyle{dtreeleaf}=[draw=black!60,minimum width=1cm,minimum height=0.4cm,rectangle,fill=blue!50!white]
\tikzset{every loop/.style={looseness=7}}
\tikzset{
	gluon/.style={decorate,draw=black,
		decoration={coil,amplitude=1pt, segment length=5pt}}
}
\tikzset{
	gluon1/.style={decorate,draw=black,
		decoration={coil,amplitude=3pt, segment length=3pt}}
}
\tikzset{
	gluonew/.style={decorate,draw=black,
		decoration={coil,amplitude=1pt, segment length=2pt}}
}

\tikzset{bicolor/.style args={#1 and #2 and #3}{
		path picture={
			\tikzset{rounded corners=0}
			\fill [#1] (path picture bounding box.south west)
			rectangle
			($(path picture  bounding box.north west)!#3!(path picture bounding
			box.north east)$);
			\fill [#2]
			($(path picture bounding box.south west)!#3!(path picture bounding
			box.south east)$)
			rectangle (path picture bounding box.north east);
}}}

\tikzset{tricolor/.style args={#1 and #2 and #3 and #4 and #5}{
		path picture={
			\tikzset{rounded corners=0}
			\fill [#1] (path picture bounding box.south west)
			rectangle
			($(path picture  bounding box.north west)!#4!(path picture bounding
			box.north east)$);
			\fill [#2]
			($(path picture bounding box.south west)!#4!(path picture bounding
			box.south east)$)
			rectangle
			($(path picture  bounding box.north west)!#5!(path picture bounding
			box.north east)$);
			\fill [#3]
			($(path picture bounding box.south west)!#5!(path picture bounding
			box.south east)$)
			rectangle (path picture bounding box.north east);
}}}

\usepackage[many]{tcolorbox}
\tcbuselibrary{listings,skins}

\lstdefinestyle{mystyle}{
  xleftmargin=0pt,
   basicstyle={\footnotesize\ttfamily},
   aboveskip=3mm,
   belowskip=3mm,
   keywordstyle=\bfseries,
   showstringspaces=false,
  escapechar=?,
  language=Java
}
\definecolor{code_indent}{HTML}{CCCCCC}

\newtcblisting{mylisting}[2][]{
  arc=0pt, outer arc=0pt,
  listing only,
  listing style=mystyle,
  title={\large #2},
  #1
}

 \definecolor{dkgreen}{rgb}{0,0.6,0}
 \definecolor{gray}{rgb}{0.5,0.5,0.5}
 \definecolor{mauve}{rgb}{0.58,0,0.82}

\definecolor{cadmiumgreen}{rgb}{0.0, 0.42, 0.24}
\definecolor{verde}{rgb}{0.25,0.5,0.35}
\definecolor{jpurple}{rgb}{0.5,0,0.35}
\definecolor{darkgreen}{rgb}{0.0, 0.2, 0.13}

\usepackage[ruled,vlined,linesnumbered,lined]{algorithm2e}
\usepackage{algpseudocode}

\usepackage{mathtools,xparse}

\usepackage{tabu}
\usepackage{multirow}

\usepackage{pifont}
\usepackage{enumerate}
\usepackage[shortlabels]{enumitem}

\usepackage{pgfplotstable}
\usepackage{pgfplots}

\usepackage[noframe]{showframe}
\usepackage{framed}

 {\endMakeFramed}
 \definecolor{shadecolor}{gray}{0.85}

\definecolor{bgblue}{RGB}{245,243,253}
\definecolor{ttblue}{RGB}{91,194,224}

\mdfdefinestyle{mystyle}{%
  rightline=true,
  innerleftmargin=10,
  innerrightmargin=10,
  outerlinewidth=3pt,
  topline=false,
  rightline=false,
  bottomline=false,
  skipabove=\topsep,
  skipbelow=false%\topsep
}

\newtcolorbox{myboxi}[1][]{
  breakable,
  title=#1,
  colback=white,
  colbacktitle=white,
  coltitle=black,
  fonttitle=\bfseries,
  bottomrule=0pt,
  toprule=0pt,
  leftrule=3pt,
  rightrule=3pt,
  titlerule=0pt,
  arc=0pt,
  outer arc=0pt,
  colframe=black!50,
}

\newtcolorbox{myboxii}[1][style=mystyle]{
  breakable,
  freelance,
  colback=white,
  colbacktitle=white,
  coltitle=black,
  fonttitle=\bfseries,
  bottomrule=0pt,
  boxrule=0pt,
  colframe=white,
  after skip=0pt,
  overlay unbroken and first={
    \draw[white!75!black,line width=3pt]
    ([yshift=-9pt]frame.north west) --
    ([yshift=9pt]frame.south west);
  },
}
\begin{document}

\title{Predicting Fairness of ML Software Configurations}

\author{Salvador Robles Herrera}
\orcid{0009-0003-2225-6779}
\affiliation{%
  \institution{University of Texas at El Paso}
  \city{El Paso}
  \country{USA}
}
\email{sroblesher1@miners.utep.edu}

\author{Verya Monjezi}
\orcid{0000-0001-6796-5948}
\affiliation{%
  \institution{University of Texas at El Paso}
  \city{El Paso}
  \country{USA}
}
\email{vmonjezi@miners.utep.edu}

\author{Vladik Kreinovich}
\orcid{0000-0002-1244-1650}
\affiliation{%
  \institution{University of Texas at El Paso}
  \city{El Paso}
  \country{USA}
}
\email{vladik@utep.edu}

\author{Ashutosh Trivedi}
\orcid{0000-0001-9346-0126}
\affiliation{%
  \institution{University of Colorado Boulder}
  \city{Boulder}
  \country{USA}
}
\email{ashutosh.trivedi@colorado.edu}

\author{Saeid Tizpaz-Niari}
\orcid{0000-0002-1375-3154}
\affiliation{%
  \institution{University of Texas at El Paso}
  \city{El Paso}
  \country{USA}
}
\email{saeid@utep.edu}

% This is the link to Collab for reproducing the results:
% https://colab.research.google.com/drive/1mr5qh11CbQbRWNwPkDTOvd2mDOj17F56#scrollTo=p3Gvm2G3fiWJ

\begin{abstract}
This paper investigates the relationships between 
hyperparameters of machine learning and fairness. Data-driven
solutions are increasingly used in critical socio-technical applications
where ensuring fairness is important.
Rather than explicitly encoding decision logic via control and 
data structures, the ML developers provide input data,
perform some pre-processing, choose ML algorithms, and tune hyperparameters (HPs) to infer
a program that encodes the decision logic.
Prior works report that the selection of HPs
can significantly influence fairness. 
However, tuning HPs
to find an ideal trade-off between accuracy, precision, and fairness
has remained an expensive and tedious task. 
\emph{Can we predict the fairness of HP configuration for a given dataset?} 
\emph{Are the predictions robust to distribution shifts?}

We focus on group fairness notions and investigate the HP space of 5 training algorithms.  We first find that tree regressors and XGBoots significantly outperformed deep neural networks and support vector machines in accurately predicting the fairness of HPs. When predicting the fairness of ML hyperparameters under temporal distribution shift, the tree regressors outperform the other algorithms with reasonable accuracy. However, the precision depends on the ML training algorithm, dataset, and protected attributes. For example, the tree regressor model was robust for training data shift from 2014 to 2018 on logistic regression and discriminant analysis HPs with sex as the protected attribute; but not for race and other training algorithms. Our method provides a sound framework to efficiently perform fine-tuning of ML training algorithms and understand the relationships between HPs and fairness. 
\end{abstract}

  

\maketitle

\vspace{-1.5 em}
\section{Introduction}
\label{sec:intro}
Artificial intelligence (AI) has become an integral part of modern software solutions that assist
in socio-economic and legal-critical decision-making processes such as releasing patients~\cite{healthcareapp},
identifying loan defaults~\cite{Home-Credit-Default-Risk},
and detecting tax evasion~\cite{nyt2023taxaudit}.
Since AI-driven software solutions are derived from prior experiences, it is not
surprising that they encode social biases in their decision logic that may
discriminate against vulnerable communities and violate ethical and legal requirements. 

Unfortunately, discriminatory software is a major threat to the trustworthiness of
automated decision-support systems, and fairness defects are plentiful in real systems.
Parole decision-making software was found to harm black and Hispanic defendants by falsely predicting a higher risk of recidivism than for non-Hispanic white defendants~\cite{compas-article}; Amazon's hiring algorithm disproportionately rejected more female applicants than male applicants~\cite{Amazon-same-day-delivery};
and data-driven auditing algorithm selected black taxpayers with earned income tax credit claims (EITC)
at much higher rates than other racial groups for an audit~\cite{nyt2023taxaudit,10173919}.
As evidenced by these examples, resulting software may particularly disadvantage minorities
and protected groups and be found non-compliant with law such as the US Civil
Rights Act~\cite{blumrosen1978wage}. Therefore,
helping programmers detect and mitigate fairness bugs
in social-critical data-driven software systems is crucial to
ensure inclusion in our modern, increasingly digital society.

Recently, the software engineering community has made concentrated efforts to improve fairness of data-driven software~\cite{zhang2021ignorance,10.1145/3368089.3409697,chakraborty2019software,tizpaz2022fairness,10.1109/ICSE48619.2023.00133,10.1145/3468264.3468536,tizpaz2024gauge}. One area of focus is on establishing best practices for
fairness of training data-driven software that involves \emph{pre-processing}~\cite{10.1145/3468264.3468536},
\emph{feature selection}~\cite{zhang2021ignorance}, and \emph{hyperparameter selection}~\cite{chakraborty2019software,tizpaz2022fairness,nguyenmake}. Specifically, hyperparameter tuning (or optimization) is the process of tweaking the configuration
space of ML training algorithms to infer an ML model, based on some
validation dataset, that provides an ideal trade-off between complexity and performance.
Some prominent examples of hyperparameters include {\tt l1} vs.\ {\tt l2} loss
function in support vector machines, the maximum depth of a decision tree,
and the number of layers/neurons in neural networks.

Chakraborty et al.~\cite{chakraborty2019software} were the first to observe the influence of
hyperparameter (HP) configurations and fairness. \textsc{Fairway}~\cite{10.1145/3368089.3409697}
searched the space of HPs to mitigate software discrimination.
\textsc{Parfait-ML}~\cite{tizpaz2022fairness} found and localized discriminatory HPs
in the prevalent ML algorithms. While the critical role of HPs on fairness
is evident, exploring the space of HPs to develop fair ML software is time-consuming and expensive due to the need to train the entire model for any given configuration.
\emph{This paper investigates whether a relationship between HPs of different ML algorithms and fairness can be learned via ML regression methods.} If feasible, the models can be used
to predict whether a HP meets fairness requirements before performing a (full) training cycle. Thus, it can significantly reduce the required computations of hyperparameter tuning.

As new data is typically acquired on an annual or periodic basis and ML system evolves over time, it becomes imperative to monitor how the fairness of a specific HP configuration changes over time. Hence, this paper also analyzes the robustness of fair configuration of HPs under temporal distribution shifts. We emphasize on time distribution shifts and investigate when and how robustness is affected or not. We pose the following research questions:

\begin{itemize}
    \item \emph{Can we accurately predict group fairness of ML hyperparameters for a given training dataset and protected attribute? what is the performance of neural methods, support vectors, and trees as the prediction methods?}
    \item \emph{Can we predict the fairness of ML hyperparameters under the temporal distribution shift of test data? which types of ML algorithms are more robust to the shift?}
    % \item \emph{To what extent can we accurately predict group fairness of ML hyperparameters over a combination of multiple datasets with the same protected attribute?}
    % \item \emph{Can we predict fairness of ML hyperparameters for a protected attribute using the ML prediction
    % model of another protected attribute but over the same dataset?}
\end{itemize}

To answer these research questions, we perform experiments over the HP space of five popular ML algorithms (Decision Tree Classifier, Support Vector Machine, Logistic Regression Classifier, Random Forest, and
Discriminant Analysis) trained over four socially-critical datasets (Adult Census,
Compas Recidivism, Default Credit, and Bank Marketing). To measure fairness,
we use the average odd difference (AOD) which is the average of differences between the true positive rates and the false positive rates of two protected groups (e.g., male vs. female).

We first use an evolutionary search algorithm to explore various HP
configurations and record the fairness of resulting models (over some validation datasets)~\cite{tizpaz2022fairness}. 
Then, we feed the collected datasets of HPs (features) and fairness (outcome)
to four ML regression methods (Deep Neural Network, Support Vector Regressor, Tree Regressors, and XGBoost) to learn a function from HPs to fairness. 
To measure the accuracy of prediction models, we used four metrics including
the coefficient of determination (known as $R^2$ score). 

Over a fixed training dataset, we observe that Tree Regressors and XGBoost outperform other algorithms in accurately predicting AOD fairness of ML hyperparameters for all five algorithms. In 40\% of cases, they achieved
$R^2 \geq 0.95$, and only in 6.7\% of cases, the algorithms do not achieve a reliable accuracy (an $R^2 \leq 0.5$). Under the temporal distribution shifts,
we observe that the precision depends on the training algorithm, dataset, and the protected attribute. When there is one year shift (e.g., trained over the income census 2014 and predicting for 2015), Tree Regressor and XGBoost achieved high accuracy in 20\% of benchmarks. We observed that these cases are related to
the HP space of Logistic Regression and Discriminant Analysis with sex as the protected attribute; and the precision is significantly degraded for other training algorithms and protected attributes like race. 

% When the fairness models obtained from a particular release of a dataset, are used to predict the fairness of previous or consequent release (one-year shift) of the same datasets, We discovered that when these models, specifically Tree Regressor and XGBoost, are used to predict fairness in datasets with a one-year temporal shift, they outperform other algorithms. In 20\% of these cases, our models achieved excellent accuracy. However, when the temporal shift extends to three years, we noticed a slight decline in the models' effectiveness. Despite this, they still maintained reasonable accuracy in 20\% of the scenarios for predicting fairness over this longer time frame. Notably, the robustness of our fairness models appeared to be influenced by the type of ML system and the protected attribute in the dataset. For instance, Tree Regressor and XGBoost showed remarkable robustness against temporal distribution shifts when predicting fairness in  hyperparameters, particularly concerning the Sex attribute. This convinces us that 
% our approach can model the fairness of ML hyperparameters when facing a three-year distribution shift in the test data. 

In summary, our experiences provide a sound and robust framework to systematically
examine the influence of hyperparameters over fairness. Our vision of usage
describes how collecting and training fairness characteristics of hyperparameters can help reduce the cost of training data-driven software solutions by avoiding biased configurations and leveraging promising hyperparameters. We also observe the difficulty of making such predictions in general, and point out circumstances for successful usages and challenges for future research.

\section{Background}
\label{sec:background}
We consider \textit{regression problem} tasks where the \textit{target} variable
is a group fairness metric, and \textit{features} are the hyperparameter variables
of ML training algorithms. The values of hyperparameter variables and target fairness variable are collected after running an evolutionary hyperparameter optimization algorithm. 

\vspace{0.25em}
\noindent \textbf{Fairness Notion.}
Fairness definitions are either concerned with individual fairness or group fairness.
Examples of individual fairness is fairness through awareness (FTA)~\cite{dwork2012fairness}
that requires two \textit{individuals} with similar
non-protected attributes are treated similarly, regardless of their background. 
\textit{Group fairness} requires the statistics of ML outcomes for different
\emph{protected groups} remain similar~\cite{hardt2016equality}.
There are multiple metrics to measure group fairness.
Among them, \textit{equal opportunity difference} (EOD) measures
the difference between the true positive rates (TPR) of two protected groups.
Similarly, \textit{average odd difference} (AOD) is the average of differences
between the true positive rates (TPR) and the false positive rates (FPR) of
two protected
groups~\cite{bellamy2019ai,10.1145/3368089.3409697,zhang2021ignorance,tizpaz2022fairness}.

\vspace{0.25em}
\noindent \textbf{ML training process.}
In the data-driven paradigm, the ML programmers often provide input data
and build an ML model using a programming interface~\cite{10.1145/3395363.3404540}.
The interface interacts with the core ML algorithms and
builds different ML models. At the heart of the training process,
tweaking HPs is particularly challenging since they cannot be estimated
from the input data, and there is no analytical formula to calculate its values~\cite{kuhn2013applied}. Examples of HPs are tolerance of optimization in SVMs, maximum features to search
in random forests, and minimum samples in leaf nodes of decision trees. We distinguish
ML training HPs from \textit{model parameters} that
are inferred automatically after training.

\vspace{0.25em}
\noindent \textbf{Related Work.}
 There are multiple works that studied
the influence of HPs on fairness~\cite{chakraborty2019software,10.1145/3368089.3409697,tizpaz2022fairness}.
Evolutionary algorithms have been significantly used to optimize the HP
search in the training process~\cite{stanley2002evolving,miikkulainen2019evolving,real2017large,xie2017genetic}.
However, there are a few tools to perform Pareto-optimal search in the
space of HPs to infer an idea trade-off between fairness (e.g., EOD and AOD) 
and accuracy (e.g., F1 socre). Parfait-ML~\cite{tizpaz2022fairness} is a gray-box evolutionary search algorithm that explores the
training space of ML algorithms to manifest HPs that minimize/maximize
fairness within an acceptable range of accuracy requirements. Given a training algorithm
and a fairness-sensitive dataset, Parfait-ML~\cite{tizpaz2022fairness} generates a set of
HPs that characterize the Pareto-dominant curves of fairness and unfairness
based on a group fairness metric such as EOD and AOD.
All of these works focus on either testing the space of HPs to find biased configurations or mitigating biased ML models by carefully picking HPs. Instead, we focus on learning the fairness characteristics of ML HPs in order to improve the
efficiency and fairness of ML software development.

\section{Fairness of ML Hyperparameters}
\label{sec:problem}
While the relationships between ML hyperparameters and functional metrics
such as overall accuracy, precision, and F1 scores have been well studied,
the presence of fairness makes the hyperparameter tuning challenging.
Our goal is to study whether a relationship between hyperparameters and
fairness can be learned via regression ML methods. 

\vspace{0.25em}
\noindent{\bf The ML Paradigm.}
Machine learning software often deploys mature, off-the-shelf ML libraries to
learn various models from training data. We can succinctly view a training problem as
the problem of identifying a mapping $M: \Xx \to \Yy$ from a set $\Xx$ of inputs to a set $\Yy$ of outputs by
learning from a fixed dataset $\Dd = \set{({\bf x_i}, {\bf y_i})}_{i=1}^N$ so that
$M$ generalizes well to previously unseen situations.

The training process includes the set $\Hh$ of hyperparameters that let the
users define the hypothesis class for the learning tasks.
The data-driven programs sift through the given dataset $\Dd$ to learn
an ``optimal'' value $\theta \in \Theta_h$ and thus compute the learning model
$M_h(\theta|\Dd): \Xx \to \Yy$ automatically.
When $\Dd$ and $\theta$ are clear from the context, we write $M_h$ for the
resulting model.

The fitness of a hyperparameter $h \in \Hh$ is evaluated by computing the
accuracy (ratio of correct results) of the model $M_h$ on a validation
dataset $\Dd_*$. We denote the accuracy of a model $M$ over $\Dd_*$ as
$ACC^M$. The dataset $\Dd_*$ is typically distinct from $\Dd$ but assumed to be sampled
from the same distribution. In the fairness-sensitive applications,
we also have fairness requirements. Let predicate $\pi: \Xx \to \set{0, 1}$
be the input variables characterizing the protected status of a data point
${\bf x}$ (e.g., race, sex, or age).
Without loss of generality, we assume there are only two protected groups:
a group with $\pi({\bf x}) = 0$ and a group with $\pi({\bf x}) = 1$.
We also assume that the predicate $\phi: \Yy \to \set{0, 1}$ over the output variables characterizes
a favorable outcome (e.g., low re-offend risk) with $\phi({{\bf y}}) = 1$.

Given $\Dd_*$ and $M: \Xx \to \Yy$, we define
true-positive rate (TPR) and false-positive rate (FPR) for the protected group
$i \in \set{0, 1}$ as
% \vspace{0.5em}
\begin{eqnarray*}
  TPR^M(i) &=& \frac{\big|\set{({\bf x}, {\bf y}) \in \Dd_* : \pi({\bf x}) = i,  \phi(M({\bf x})) = 1, \phi({\bf y}) = 1}\big|}
       {\big|\set{({\bf x}, {\bf y}) \in \Dd_* : \pi({\bf x}) = i}\big|}\\
  FPR^M(i) &=& \frac{\big|\set{({\bf x}, {\bf y}) \in \Dd_* : \pi({\bf x}) =
       i, \phi(M({\bf x})) = 1, \phi({\bf y}) = 0}\big|}
       {\big|\set{({\bf x}, {\bf y}) \in \Dd_* : \pi({\bf x}) = i}\big|}.
\end{eqnarray*}
We use a prevalent
notions of group fairness~\cite{bellamy2019ai,10.1145/3368089.3409697,zhang2021ignorance}:
% \srsays{Why does the citation appear like "3??" instead of "3"!}
% \stsays{Good catch: fixed it!}
\textit{equal opportunity difference} (EOD) of $M$ against $\Dd_*$
between two groups is
% \vspace{0.1em}
% \[
$EOD^M = \big|TPR^M(0) - TPR^M(1) \big|$,
% \]
and \textit{average odd difference} (AOD) of $M$ against $\Dd_*$ between two groups is
% \vspace{0.1em}
\[
AOD^M = \frac{|TPR^M(0)-TPR^M(1)| + |FPR^M(0)-FPR^M(1)|}{2}.
\]
We notice that $0 \leq EOD^M \in \R \leq 1$ and $0 \leq AOD^M \in \R \leq 1$,
and higher values of $EOD^M$ and $AOD^M$ indicate low fairness. 

The key idea behind hyperparameter optimization methods~\cite{tizpaz2022fairness,10.1145/3368089.3409697}
for fairness and accuracy is to search the space of hyperparameters and find hyperparameters that maximize
fairness, while making sure that the accuracy does not degrade below a given threshold.
To understand the effects of hyperparameters on fairness, these methods also
allow us to find hyperparameters that minimize fairness. The both sets of hyperparameters
that manifest maximum and minimum fairness values provide us with representative samples to
systematically investigate the relationships between hyperparameters and fairness.

% A {\it fairness trace} $\Ff$ of an ML training algorithm $M$ is a tuple $(\Dd_i, h_i^j, p, F_{i,j,p})$
% where $\set{\Dd_1, \ldots, \Dd_d}$ is a set of training/validation datasets,
% $h_i^j \in \Hh$ is hyperparameter value,
% $p$ is a predicate over protected attributes,
% and $F_{i,j,p}$ is the group fairness metric of ML model $M_h$ for the predicate $p$
% after training $M$ with the dataset $\Dd_i$ using the hyperparameter $h_i^j$.

A {\it fairness trace} $\Ff$ of an ML training algorithm $M$ is characterized by the tuple $(\Dd, \Hh, \Pi, F)$
where $\Dd$ is the training dataset,
$\Hh$ is the hyperparameter,
$\Pi$ is a predicate over protected attributes,
and $F \in \set{EOD,AOD}$ is the group fairness metric of ML model $M_h^D$ when trained
over a dataset $D \in \Dd$ with the hyperparameter $h \in \Hh$ and
the protected attribute predicate $p \in \Pi$.

% \begin{definition}
%     \label{def-problem}
    \begin{tcolorbox}[boxrule=1pt,left=1pt,right=1pt,top=1pt,bottom=1pt]
     Given a set of fairness training set $\Ff$, 
     our goal is to infer a function $A(\omega|D, p): \Hh \to [0,1]$ with different hypothesis classes
     of $\omega$ (e.g., linear, tree, and neural network regressors) over fairness training traces $\Ff$
     that generalizes well to the testing traces $\Ff_*$,  i.e., learn fairness as a function of hyperparameters for a fixed dataset $D \in \Dd$ and protected predicate $p \in \Pi$. We call $A$ and $A_w$ HP prediction algorithm and model, respectively. 
    \end{tcolorbox}
% \end{definition}

From the practical point of view, it is not only important that fairness of hyperparameters
of ML training algorithms can be predicted for a specific task at a specific time;
but also whether a similar relationship can be learned 
under distribution shifts (e.g., under covariate shift~\cite{kull2014patterns,varshney2021trustworthy} when the feature distributions are different). One specific type of shifts is temporal distribution shifts where our goal is to evaluate whether the prediction model is robust to the temporal changes.
In other words, we study the accuracy of $A_{D^0 \to D^k}$, that is to evaluate the ML prediction algorithm $A(\omega|D^0, p)$, trained with the base dataset at the time unit $0$, with the dataset $D^k$, sampled after $k$ units of time (e.g, after $k$ year).

% \begin{definition}[Learning Fairness of Hyperparameters]
%     \label{def-problem}
%     \begin{tcolorbox}[boxrule=1pt,left=1pt,right=1pt,top=1pt,bottom=1pt]
% Given a set of fairness traces $\Ff^M = (\Dd_i, h_i^j, p, F_{i,j,p})$ (split into
% training and testing subsets) and several machine learning tools $A$, to study: i) for each tool $A$, database $i$, and a protected predicate $p$, whether it is possible to train the machine learning tool $A$ so that this trained tool $A_{i,p}$ would be able, given hyperparameter $h^i_j$, to accurately estimate the corresponding average odd difference $F_{i,j,p}$: $A_{i, p}(h^i_j)\approx F_{i,j,p}$;
% ii) whether the tool $A_{i,p}$ trained on one dataset $i$ will work on another dataset $i'$, i.e., whether $A_{i,p}(h^{i'}_j)\approx F_{i,j,p}$; and
% ii) whether the tool $A_{i,p}$ trained on one protected predicate $p$ will work on another protected predicate $p'$, i.e., whether $A_{i,p}(h^{i}_j)\approx F_{i,j,p'}$.

%     \end{tcolorbox}
% \end{definition}

\section{Learning Fairness Functions}
\label{sec:approach}
Our goal is to infer $\omega$ of HP
prediction algorithms $A$ to accurately predict the fairness of ML algorithm
configurations. There are many different ML algorithms that can be used to solve the inference
problem. After some initial experimentation,
we focus on four types of $\omega$ with the following prediction algorithms:

% We initially consider deep neural networks, linear regressions, support vector regression, polynomial regression, and RANSAC regressor. However, after initial experimentation, we excluded the decision tree regressor,
% polynomial regression, and RANSAC regressor due to unacceptable performance.

% \begin{itemize}
   \noindent \textit{1) Deep Neural Network} (DNN). We primarily work with feedforward neural networks with ReLU
    activation units. A feedforward neural network is characterized by its number of hidden layers,
    the input and output dimensions, and the number of neurons in each layer. Each hidden layer $i$ implements an affine
    mapping $T_i$ that takes outputs of the previous layer and applies a linear function using its weights along with
    ReLU activation function ($\sigma$). The function $f_\Nn: \Real^{w_0} \to \Real^{w_{N+1}}$ implemented by
    DNN $\Nn$ is
%    \[
    $A_\omega = T_{k+1} \circ \sigma \circ T_k \circ \sigma \circ \cdots T_2 \circ \sigma
    \circ T_1.$
%    \]

  \noindent \textit{2) Support Vector Regression} (SVR). 
  SVR follows the same principles as a support vector machine (SVM) that aims to infer the parameters of a hyperplane that separate
  data points of different class labels with an error margin. Since $A$ is a regression task, we change the objective to
  infer a hyperplane that can predict target outcomes within a tolerance margin~\cite{drucker1996support}, i.e.,
  to minimize $\frac{1}{2}||\omega^2||$ subject to $|F - \omega.\Hh - b | \leq \epsilon$.
  
  \noindent \textit{3) Tree Regressor} (TR). The algorithm uses an ensemble method where it trains a number of regression trees in parallel~\cite{loh2011classification}.
  At the inference time, the output is the mean of predicted outcomes from each regression tree. 
  
  \noindent \textit{4) XGBoost} (XGB). This algorithm is similar to the Random Forest Regressor, but rather than an ensemble method,
  it uses tree boosting techniques~\cite{friedman2001greedy} that start from a single leaf and iteratively
  add branches to the tree until the performance is not improving. XGBoost~\cite{chen2016xgboost} is a special tree boosting algorithm that scales the training process via sparse data analysis and approximating training. XGBoost
  has been shown to significantly outperform other tree boosting techniques~\cite{bentejac2021comparative}. 
% \end{itemize}

Our goal is to compare the performance of different hypothesis classes of HP prediction algorithms.
If feasible, the models can be used as a part of data-driven software developments to predict the fairness of a selected HP without requiring a full training cycle. \emph{In summary, our goal is to study the feasibility
of training regression models to predict the fairness of ML hyperparameters in two levels:
i) fixed dataset over a protected attribute and ii) temporal distribution shifts between training
and test data.}

\section{Experiments}
\label{sec:experiments}
We first give background on ML training algorithms and their hyperparameters. 
Then, we overview training datasets used to infer ML models and evaluate their
fairness. Subsequently, we introduce ML prediction algorithms (regressors)
that are used to learn the fairness models of hyperparameters. Finally,
we discuss research questions and perform experiments to evaluate the performance
of different algorithms under different circumstances. 

\vspace{0.25 em}
\noindent \textbf{ML Training Algorithms.}
We consider $5$ different ML training algorithms provided by \textit{scikit-learn}, a popular ML library.
The following is an overview of these algorithms: 
% \begin{itemize}

\noindent \textit{A) Decision Tree} (DT) algorithm~\cite{Decision-Tree} is a supervised learning algorithm to infer tree classifiers. 
% If we visualize the models, they look like a tree, where each node represents a variable and each edge shows a split over its value to visit a left or right sub-tree.
During training, the DT algorithms use information theoretic measures, known as impurity, to decide which feature (node) and values (edges) to split upon. The final level of the tree contains leaf nodes that decide the outcome labels. We consider the following HPs:  max\_depth, min\_samples\_split, min\_samples\_leaf, and min\_weight\_fraction\_leaf
(\textbf{numerical}); and Criterion, splitter, and max\_features (\textbf{Categorical}). 

\noindent \textit{B) Logistic Regression} (LR) infers the parameters of sigmoid functions
to map data into a likelihood score of receiving a particular label. 
We use an implementation from scikit-learn~\cite{logistic-regression} that has the following
hyperparameters: tol, C, intercept\_scaling, max\_iteration, and l1\_ratio (\textbf{numerical});
and solver, penalty norm, dual\_prime formulation, fit\_intercept, and multi\_class (\textbf{Categorical}).

\noindent \textit{C) Support Vector Machine} (SVM) learns a hyperplane of lower dimensions to separate
the data of high-dimensional spaces on the basis of their class labels. 
% An SVM algorithm performs
% an optimization to choose the best hyperplane, which is the one that represents the largest 
% separation between two classes. 
The scikit-learn implementations of SVM algorithms~\cite{SVM}
include the following hyperparameters: tol, C, and intercept\_scaling (\textbf{numerical});
and penalty, loss, degree, fit\_intercept, and class\_weight (\textbf{Categorical}).

\noindent \textit{D) Random Forest} (RF) is an ensemble learning algorithm that fits a number
of trees and takes the expected outcomes of these trees to make a prediction. We 
used an implementation from scikit-learn with the following hyperparameters:
max\_depth, min\_samples\_split, min\_samples\_leaf, min\_weight\_fraction\_leaf, n\_estimators, 
and max\_samples (\textbf{numerical}); and criterion, max\_features, oob\_score, and warm\_start (\textbf{Categorical}).

\noindent \textit{E) Discriminant Analysis} (DA) defines a smooth classifier where the data fits
a Gaussian prior to inferring a posterior for each class label~\cite{Discriminant-Analysis}.
% The distribution of posterior is used to predict the class of test data.
The scikit-learn
implementations of DA algorithms include the following hyperparameters: tol, and reg\_param
(\textbf{numerical}); and linear(0)\_quadratic(1), solve\_Linear, Shrinkage\_Linear, component,
store\_covariance, and type\_dataset (\textbf{Categorical}). 

% \end{itemize}

% Each algorithm is used on different datasets, this are the datasets (each containing its different protected attributes):
% \begin{itemize}
% \item Census: This is the adult census income dataset, the main purpose of this data is to predict whether an individual earns more than \$50k annually given some features. Some of the attributes are Education level, Age, Gender, Occupation, etc.
% \item Credit: This dataset contains credit data of individuals. Some of the attributes are ID, Gender, Education, Marriage, Age, etc.
% \item Bank: This dataset
% \end{itemize}

{\footnotesize
\begin{table*}[!t]
\caption{Fairness-sensitive datasets used as the training of ML algorithms.}
\centering
\resizebox{0.8\textwidth}{!}{
\begin{tabu}{|l|l|l|ll|ll|}
  \hline
  \multirow{2}{*}{\textbf{Dataset}} & \multirow{2}{*}{\textbf{|Instances|}} & \multirow{2}{*}{\textbf{|Features|}} & \multicolumn{2}{c|}{\textbf{Protected Groups}} & \multicolumn{2}{c|}{\textbf{Outcome Label}} \\
   &  & & \textit{Group1} & \textit{Group2} & \textit{Label 1} & \textit{Label 0}  \\
  \hline
  Adult \textit{Census} & \multirow{2}{*}{$48,842$} & \multirow{2}{*}{$14$} & Sex-Male & Sex-Female & \multirow{2}{*}{High Income} & \multirow{2}{*}{Low Income} \\ \cline{4-5}
  Income (Ce) &  &    &  Race-White & Race-Non White  &   &    \\
  \hline
  \multirow{2}{*}{\textit{Compas} Software (Co)}  & \multirow{2}{*}{$7,214$} & \multirow{2}{*}{$28$} & Sex-Male & Sex-Female & \multirow{2}{*}{Did not Reoffend} & \multirow{2}{*}{Reoffend} \\  \cline{4-5}
  &  &    &  Race-Caucasian & Race-Non Caucasian  &   &    \\
  \hline
  German \textit{Credit} (Cr)  & $1,000$ & $20$ & Sex-Male & Sex-Female & Good Credit & Bad Credit \\
  \hline
  \textit{Bank} Marketing (Ba)  & $45,211$ & $17$ & Age-Young & Age-Old & Subscriber & Non-subscriber \\
  \hline
\end{tabu}
}
\label{table:dataset}
\end{table*}
}

\vspace{0.25 em}
\noindent \textbf{Fairness-Sensitive Training Datasets.}
We consider $4$ fairness-critical datasets from the literature of
ML fairness to infer ML classifiers using the ML training algorithms
~\cite{10.1145/3368089.3409697,FairnessTesting,10.1145/3338906.3338937,tizpaz2022fairness}.
Table~\ref{table:dataset} shows these datasets with different protected
attributes that form $6$ training tasks. 
Adult Census Income~\cite{Dua:2019-census} ($Ce$) is a binary classification dataset
that predicts whether an individual has an income over $50K$ a year. 
Sex and race are two protected attributes for this dataset. 
COMPAS Software ($Co$)~\cite{compas-dataset} is a dataset in the criminal justice domain
that classifies defendants into high- or low-risks for re-offending (i.e., risk of recidivism). We
consider sex and race as the protected attributes. German credit data ($Cr$)~\cite{Dua:2019-credit}
predicts whether an individual might have bad credit in assessing a loan default. 
%\srsays{Last sentence. Should it be "for receiving a loan default"?}.
%\stsays{loan default is a bad thing}
Sex is the protected attribute. Bank Marketing ~\cite{Dua:2019-bank} is a dataset
in the advertisement applications to predict the likelihood that an individual will be
a subscriber.

\vspace{0.25 em}
\noindent \textbf{Training ML prediction algorithms.}
We use \textsc{Parfait-ML}~\cite{tizpaz2022fairness} to generate a dataset of HP
values (as feature variables) and AOD fairness values (as the target variable). Given a ML training algorithm (e.g., decision tree)
and a fairness-sensitive dataset (e.g., Adult Census); we run \textsc{Parfait-ML} for four hours
and record all the HPs as well as the AOD fairness of the corresponding ML model. 
Given the HP traces of each ML training algorithm and fairness-sensitive training datasets;
we leverage the ML prediction algorithms to train ML regressors that can accurately predict
the fairness of a HP configuration.
% To train those algorithms, we consider:
To train those models, we did standard hyperparamter tuning to obtain the following predictive models: 

% \begin{itemize}
  \noindent \textit{1. Deep Neural Network} (DNN). We consider prevalent deep neural network architecture from the fairness literature
  with $4$ hidden layers, each with 32 neurons~\cite{zhang2020white,10.1145/3510003.3510137,monjezi2023information}. 
  We implemented DNN using the TensorFlow library, version 2.14.0, where we used
  ``mean squared error'' (MSE) as the loss function and Adam for optimization. We infer the parameter of DNN after
  50 epochs (batch size of 64).
  \noindent \textit{2. Support Vector Regression} (SVR). We consider different classes of SVR for learning fairness functions. 
  After initial experimentation and fine-tuning over SVR, LinearSVR, and NuSVR algorithms; we consider NuSVR as
  the core SVR algorithm from scikit-learn, version 1.2.2, with the following configuration:
  ``rbf'' as the kernel, ``auto'' as gamma for kernel coefficient, and
  10,000 for the maximum number of iterations.
    
  \noindent \textit{3. Tree Regressor} (TR). We consider the implementation of random tree regressors from scikit-learn, version 1.2.2, where after initial experiments, we set the algorithm with 100 tree estimators each with a maximum depth of 35 nodes
  throughout the experiments. 
  
  \noindent \textit{4. XGBoost} (XGB) We consider the implementations from \textsc{XGBoost} library~\cite{chen2016xgboost}, using version 2.0.2. After initial exploration, we set the max. depth of each tree to 30. 
% \end{itemize}

\vspace{0.25 em}
\noindent \textbf{Evaluation of the HP Prediction Algorithms.}
% Given a prediction model inferred by the ML prediction algorithms to project fairness of a given
% ML hyperparameter over the fairness training traces;
Our next goal is to evaluate the accuracy of the model in predicting the AOD bias of HPs. 
The first metric is the loss function, which calculates the fitness of our predictions
compared to the actual AODs. This loss function is commonly Mean Square Error (MSE).
% Usually, MAE is preferred when the data has a lot of outliers, but in our data, we did not observe evidence of outliers.
% Therefore, we will use MSE as the loss function during training. 

To evaluate the accuracy of our models over the testing data, we use three primary metrics:
the Root Mean Square Error (RMSE), Relative Root Mean Square Error (Relative RMSE), and Coefficient
of Determination ($R^2$). Since the metrics are highly correlated, we will focus on $R^2$ as our main metric. Let $n$ be the number of (testing) samples, $\hat{AOD_i}$ be the predicted fairness from the ML prediction algorithm, ${AOD_i}$ be the observable (ground truth) fairness for the $i$-th data point, and $\overline{AOD}=\frac{\sum_{i=1}^{n}(AOD_i)}{n}$ be the average of the AOD values.
% \srsays{Instead of the small bar above a symbol to represent the average, should we put a bigger line like I did on the previous AOD average? I put overline instead of bar}
% \stsays{nice!}

% \srsays{Commented out RMSE part}
\begin{itemize}[leftmargin=*]
\begin{comment}
  \item \textbf{RMSE.} The Root Mean Square Error \vmsays{Do we need this?}
  % (RMSE) metric is sort of a percentage error metric and is derived from the MSE. The RMSE is equal to the square root of the MSE. It 
  measures the distance from the prediction values to the observed values. 
  We express this metric with the following formula (lower values mean better fit): 
\[
RMSE = \sqrt{\frac{\sum_{i=1}^{n}(\hat{AOD_i} - AOD_i)^2}{n}}
\]
\end{comment}

  \item \textbf{Relative RMSE.} The Relative Root Mean Square Error is equal to the RMSE divided by the average $\bar{AOD}$ value of our dataset (lower values show a better fit):
\[
Relative\,RMSE = \frac{1}{\overline{AOD}}*\sqrt{\frac{\sum_{i=1}^{n}(\hat{AOD_i} - AOD_i)^2}{n}} 
\]
  \item \textbf{The coefficient of determination} ($R^2$). This is the main evaluation metric in our experiments. 
  This statistic measures how well the data fits our trained ML prediction model. The following is the formula for the $R^2$ (higher values show a better fit):
    \[
    R^2 = 1-\frac{\sum_{i=1}^{n}(AOD_i - \hat{AOD_i})^2}{\sum_{i=1}^{n}(AOD_i - \overline{AOD})^2}
    \]
   This value can normally range from 0 to 1: if the $R^2$ value is 1, then we have perfectly predicted the output values whereas a lower value casts doubts on the validity of the trained models. In many practical situations, we depend on a single threshold that informs us if our $R^2$ value is reasonable. We report $R^2$ values above 0.95, 0.8, and 0.5; and consider values below 0.0 as an invalid prediction model. 
\end{itemize}

\noindent \textit{Baseline Model.} 
We also consider a basic simple prediction model to understand the precision of models.
In particular, we use the average AOD of all samples, i.e., $\overline{AOD}$ to predict
AOD of any HP values. We expected that our trained ML prediction models perform (significantly)
better than this naive predictor. 

% In order to figure out if our ML models are actually learning something valuable, we can compare them to a baseline model. What is a baseline model? It is a first simple attempt to creating a model which will be a reference point for future predictions. The advantages of a baseline model is that we can have a reference point to which we can compare our ML models. If the loss of a ML model is worse than this baseline (or fairly the same loss), then that means that the model is not learning.

% To create our baseline model we will compute the average value of the outputs $Y$, and that value will be the prediction for all the testing values we want to predict. This doesn't require any Machine Learning or any difficult process or algorithm. Is just predicting the same value regardless of the input, not something very smart to do.

\vspace{0.25 em}
\noindent \textbf{Research Questions.}
We pose the following research questions:
\begin{itemize}[leftmargin=*]
\item \textbf{RQ1.} Can we accurately predict the fairness of ML hyperparameters? 
What is the performance of different classes
of predictors?

\item \textbf{RQ2.} Can we predict the fairness of ML hyperparameters under temporal
distribution shift of dataset? 
% Which class of ML algorithms are more robust to
% temporal shift?

% \item \textbf{RQ3.} Can we accurately predict the fairness of ML hyperparameters over multiple protected
% attributes with a fixed dataset? 
% \srsays{Could it be "over multiple protected attributes with a fixed dataset"?}
\end{itemize}

\noindent \textit{Technical Details.}
We implemented our experiments using a cloud server machine with Intel(R) Xeon(R) CPU @ 2.20ghz. The system has 12.7 GB RAM,
and the disk space is 225.8 GB. Throughout this paper, we repeated experiments $10$ times each with different random seeds. We report both the average and the standard deviation over $10$ repeated experiments. \textbf{The replication package is available at} 
\href{https://figshare.com/s/311c7d5d6e406f966623}{link}.
% \srsays{Can you check the last sentence?}

% Random Seeds used: [2001, 1500, 150, 768, 345, 876, 302, 450, 112, 2000]

\begin{figure*}
    \centering
    \includegraphics[width=0.25\textwidth]{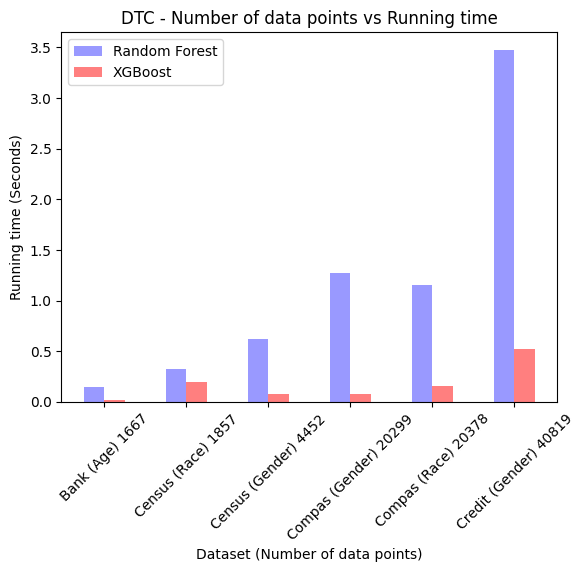}
    \hfill    
    \includegraphics[width=0.25\textwidth]{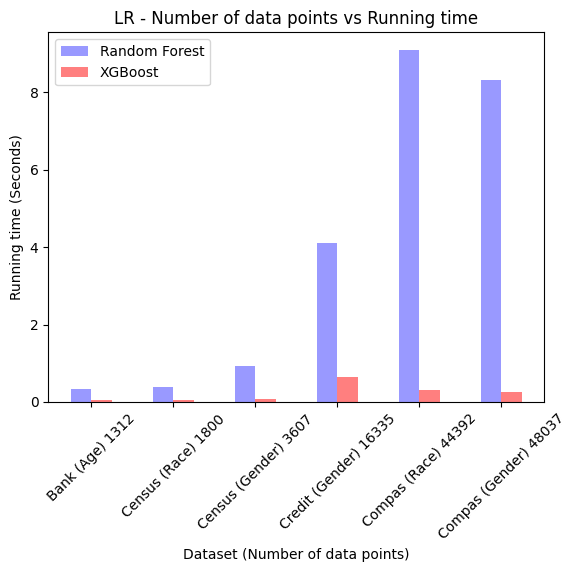}
    \hfill
    % \includegraphics[width=0.3\textwidth]{Figures/RF_complexity.png}
    % \hfill
    \includegraphics[width=0.25\textwidth]{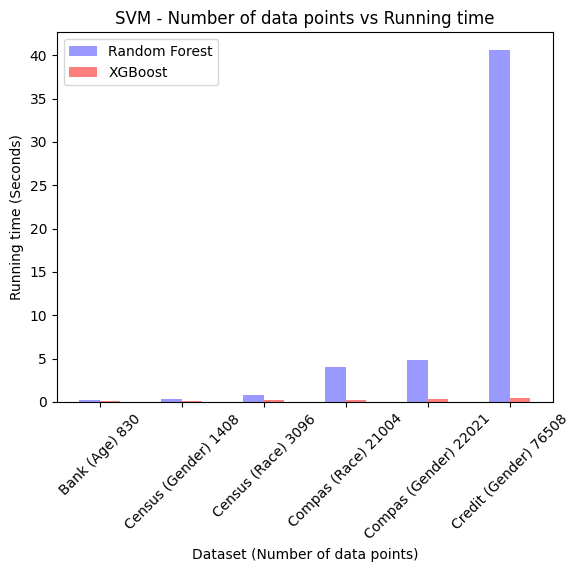}
    % \hfill
    % \includegraphics[width=0.3\textwidth]{Figures/DA_complexity.png}
    
    \caption{Computation Complexities of Tree Regressor vs. XGBoost.
    Each plot shows one ML training algorithm (Decision Tree Classifier in the left,
    Logistic Regression in the center, and Support Vector Machine in the right).
    % the x-axis shows different
    % training datasets with the size of fairness traces (as inputs to TR and XGB),
    % and the y-axis shows the computation times of training with TR vs. XGB.
    }
    \label{fig:computational-complexity}
\end{figure*}

% {\footnotesize
\begin{flushleft}
\begin{table*}[ht]
\caption{(RQ1) Comparisons between different ML prediction algorithms (DNN, SVR, TR, and XGB) in learning fairness of
hyperparameters for a given ML training algorithm, dataset, and protected attribute. The numbers
show the average of 10 repeated experiments wherein the standard deviations are reported in the parenthesis
(NV shows $R^2 \leq 0.0$).}
\centering
\resizebox{0.92\textwidth}{!}{%
\begin{tabu}{|l|l|l|l|l|l|l|l|l|l|l|l|}

\hline
\multirow{2}{4.7em}{\textbf{Algorithm}} & \multirow{2}{3.0em}{\textbf{Dataset}} & \multirow{2}{1.5em}{\textbf{Prot.}}  & \multirow{2}{2.0 em}{\textbf{Baseline}} & \multicolumn{2}{| l |}{\textsc{Deep Neural Network}} & \multicolumn{2}{| l |}{\textsc{Support Vector Regression}}  &\multicolumn{2}{| l |} {\textsc{Tree Regressor}} & \multicolumn{2}{| l |}{XGBoost}\\\cline{5-12}

 & & & & \texttt{Rel. RMSE} & $R^2$ & \texttt{Rel. RMSE} & $R^2$ & \texttt{Rel. RMSE} & $R^2$ & \texttt{Rel. RMSE}& $R^2$ \\
\hline
\multirow{6}{2.5em}{Decision Tree} & \multirow{2}{2.5em}{Census}
& sex  
& 0.444 (0.013) & 0.366 (0.018) % & 0.038 (0.002) 
& 0.316 (0.057) 
& 0.387 (0.012) % & 0.040 (0.001) 
& 0.237 (0.034) 
& 0.157 (0.014) % & 0.016 (0.001) 
& \textbf{0.875 (0.018)} 
& 0.164 (0.012) % & 0.017 (0.001) 
& \textbf{0.863 (0.016)}  \\
 & & race  
 & 0.610 (0.029) & 0.607 (0.032) % & 0.043 (0.002) 
 & 0.003 (0.099) 
 & 0.624 (0.037) % & 0.044 (0.002) % & -0.051 (0.057) 
 & NV
 & 0.233 (0.033) % & 0.016 (0.002) 
 & \textbf{0.849 (0.045)} & 0.230 (0.029) % & 0.016 (0.002)
 & \textbf{0.853 (0.042)} \\
\cline{2-12}
 & \multirow{2}{2.5em}{Compas} 
 & sex  & 0.273 (0.004) & 0.115 (0.014) % & 0.002 (0.000) 
 & 0.820 (0.050) & 0.239 (0.035) % & 0.005 (0.001) 
 & 0.219 (0.234) & 0.057 (0.003) % & 0.001 (0.000) 
 & \textbf{0.956 (0.005)} & 0.058 (0.003) % & 0.001 (0.000) 
 & \textbf{0.954 (0.005)} \\
 & 
 &  race  & 0.347 (0.003) & 0.141 (0.013) % & 0.003 (0.000) 
 & 0.832 (0.031) & 0.278 (0.012) % & 0.005 (0.000) 
 & 0.358 (0.055) & 0.068 (0.004) % & 0.001 (0.000) 
 & \textbf{0.962 (0.005)} & 0.066 (0.004) % & 0.001 (0.000) 
 & \textbf{0.964 (0.005)} \\
\cline{2-12}
& Credit 
& sex   & 2.219 (0.036) & 1.260 (0.077) % & 0.026 (0.002) 
& 0.676 (0.045) & 2.317 (0.185) % & 0.048 (0.004) % & -0.095 (0.156) 
& NV
& 0.694 (0.026) % & 0.014 (0.001) 
& \textbf{0.902 (0.008)} & 0.712 (0.031) % & 0.015 (0.001) 
& \textbf{0.897 (0.010)} \\
\cline{2-12}
& Bank & age  & 4.008 (0.658) & 4.034 (0.621) % & 0.008 (0.001) % & -0.027 (0.127) 
& NV
& 3.848 (0.675) % & 0.008 (0.001) 
& 0.076 (0.073) & 2.168 (0.533) % & 0.004 (0.001) 
& \textbf{0.706 (0.070)} & 2.039 (0.466) % & 0.004 (0.001) 
& \textbf{0.734 (0.084)} \\
\hline
\multirow{6}{2.5em}{Logistic Regression} & \multirow{2}{2.5em}{Census} 
& sex  & 0.432 (0.008) & 0.317 (0.072) % & 0.020 (0.004) 
& 0.428 (0.304) & 0.217 (0.008) % & 0.014 (0.000) 
& 0.747 (0.016) & 0.053 (0.008) % & 0.003 (0.001) 
& \textbf{0.984 (0.005)} & 0.053 (0.008) % & 0.003 (0.001) 
& \textbf{0.985 (0.005)} \\
 & & race  & 0.197 (0.010) & 0.163 (0.037) % & 0.011 (0.003) 
 & 0.285 (0.333) & 0.130 (0.016) % & 0.009 (0.001) 
 & 0.560 (0.086) & 0.044 (0.023) % & 0.003 (0.002) 
 & \textbf{0.940 (0.059)} & 0.051 (0.026) % & 0.004 (0.002) 
 & \textbf{0.920 (0.076)} \\
\cline{2-12}
 & \multirow{2}{2.5em}{Compas} 
 & sex  & 0.377 (0.002) & 0.072 (0.016) % & 0.002 (0.000) 
 & 0.962 (0.019) & 0.266 (0.005) % & 0.007 (0.000) 
 & 0.502 (0.018) & 0.008 (0.001) % & 0.000 (0.000) 
 & \textbf{1.000 (0.000)} & 0.011 (0.001) % & 0.000 (0.000) 
 & 0.999 (0.000) \\
 & &  race  & 0.372 (0.004) & 0.128 (0.039) % & 0.002 (0.001) 
 & 0.871 (0.084) & 0.318 (0.008) % & 0.006 (0.000) 
 & 0.267 (0.045) & 0.007 (0.002) % & 0.000 (0.000) 
 & \textbf{1.000 (0.000)} & 0.011 (0.002) % & 0.000 (0.000) 
 & 0.999 (0.000) \\
\cline{2-12}
& Credit & sex  & 0.400 (0.001) & 0.203 (0.022) % & 0.014 (0.002) 
& 0.741 (0.064) & 0.238 (0.005) % & 0.017 (0.000) 
& 0.645 (0.017) & 0.113 (0.004) % & 0.008 (0.000) 
& \textbf{0.920 (0.006)} & 0.116 (0.004) % & 0.008 (0.000) 
& \textbf{0.916 (0.006)} \\
\cline{2-12}
& Bank & age  & 0.679 (0.014) & 0.469 (0.049) % & 0.015 (0.001) 
& 0.515 (0.097) & 0.402 (0.021) % & 0.012 (0.001) 
& 0.646 (0.043) & 0.172 (0.018) % & 0.005 (0.001) 
& \textbf{0.935 (0.014)} & 0.182 (0.024) % & 0.006 (0.001) 
& \textbf{0.926 (0.020)} \\
\hline
\multirow{6}{2.5em}{Random Forest} & \multirow{2}{2.5em}{Census} 
& sex  & 0.220 (0.011) & 1.375 (1.335) % & 0.170 (0.166) % & -69.467 (131.431)
& NV 
& 0.210 (0.013) % & 0.026 (0.002) 
& 0.081 (0.054) & 0.055 (0.006) % & 0.007 (0.001) 
& \textbf{0.935 (0.015)} & 0.057 (0.005) % & 0.007 (0.001) 
& \textbf{0.931 (0.012)} \\
 & & race  & 0.160 (0.012) & 2.547 (1.381) % & 0.176 (0.096) % & -324.619 (303.137) 
 & NV
 & 0.162 (0.012) % & 0.011 (0.001) % & -0.030 (0.037)
 & NV
 & 0.093 (0.007) % & 0.006 (0.001) 
 & \textbf{0.657 (0.055)} & 0.095 (0.003) % & 0.007 (0.000) 
 & \textbf{0.637 (0.066)} \\
\cline{2-12}
 & \multirow{2}{2.5em}{Compas} 
 & sex  & 0.046 (0.017) & 0.074 (0.022) % & 0.001 (0.000) % & -3.816 (4.790) 
 & NV
 & 0.050 (0.009) % & 0.001 (0.000) % & -0.607 (1.066) 
 & NV
 & 0.026 (0.007) % & 0.000 (0.000) 
 & \textbf{0.567 (0.282)} & 0.022 (0.008) % & 0.000 (0.000) 
 & \textbf{0.708 (0.159)} \\
 & &  race  & 0.042 (0.007) & 0.130 (0.203) % & 0.002 (0.003) % & -27.481 (76.889) 
 & NV
 & 0.049 (0.005) % & 0.001 (0.000) % & -0.382 (0.371) 
 & NV
 & 0.019 (0.004) % & 0.000 (0.000) 
 & \textbf{0.807 (0.036)} & 0.021 (0.004) % & 0.000 (0.000) 
 & \textbf{0.739 (0.106)} \\
\cline{2-12}
& Credit & sex  & 0.625 (0.007) & 0.536 (0.086) % & 0.027 (0.004) 
& 0.246 (0.261) & 0.614 (0.008) % & 0.031 (0.000) 
& 0.034 (0.018) & 0.317 (0.004) % & 0.016 (0.000) 
& \textbf{0.743 (0.006)} & 0.338 (0.004) % & 0.017 (0.000) 
& 0.708 (0.008) \\
\cline{2-12}
& Bank & age  & 0.662 (0.118) & 27.331 (25.2) % & 0.306 (0.282) % & -2646.071 (4477.075) 
& NV
& 0.518 (0.061) % & 0.006 (0.001) 
& 0.251 (0.466) & 0.331 (0.054) % & 0.004 (0.001) 
& \textbf{0.710 (0.126)} & 0.350 (0.064) % & 0.004 (0.001) 
& \textbf{0.677 (0.139)} \\
\hline
\multirow{6}{2.5em}{SVM} & \multirow{2}{2.5em}{Census} 
& sex  & 0.828 (0.009) & 0.275 (0.022) % & 0.042 (0.003) 
& \textbf{0.889 (0.018)} & 0.293 (0.023) % & 0.045 (0.004) 
& 0.874 (0.022) & 0.244 (0.027) % & 0.037 (0.004) 
& \textbf{0.912 (0.020)} & 0.267 (0.029) % & 0.041 (0.005) 
& \textbf{0.895 (0.025)} \\
 & & race  & 0.733 (0.004) & 0.378 (0.021) % & 0.042 (0.002) 
 & \textbf{0.733 (0.029)} & 0.399 (0.025) % & 0.044 (0.003) 
 & 0.702 (0.038) & 0.355 (0.021) % & 0.039 (0.002) 
 & \textbf{0.764 (0.028)} & 0.381 (0.023) % & 0.042 (0.003) 
 & \textbf{0.729 (0.032)} \\
\cline{2-12}
 & \multirow{2}{2.5em}{Compas} 
 & sex  & 0.178 (0.001) & 0.088 (0.016) % & 0.002 (0.000) 
 & 0.751 (0.106) & 0.162 (0.029) % & 0.003 (0.001) 
 & 0.151 (0.314) & 0.066 (0.002) % & 0.001 (0.000) 
 & \textbf{0.864 (0.007)} & 0.069 (0.002) % & 0.001 (0.000) 
 & \textbf{0.850 (0.008)} \\
 & &  race  & 0.069 (0.003) & 0.063 (0.004) % & 0.001 (0.000) 
 & 0.161 (0.072) & 0.129 (0.031) % & 0.002 (0.000) % & -2.643 (1.738) 
 & NV
 & 0.049 (0.003) % & 0.001 (0.000) 
 & 0.496 (0.044) & 0.052 (0.003) % & 0.001 (0.000) 
 & 0.431 (0.050) \\
\cline{2-12}
& Credit & sex  & 0.449 (0.001) & 0.043 (0.005) % & 0.006 (0.001) 
& 0.991 (0.003) & 0.181 (0.016) % & 0.024 (0.002) 
& 0.835 (0.028) & 0.018 (0.001) % & 0.002 (0.000) 
& \textbf{0.998 (0.000)} & 0.021 (0.002) % & 0.003 (0.000) 
& \textbf{0.998 (0.000)} \\
\cline{2-12}
& Bank & age  & 0.782 (0.052) & 0.684 (0.035) % & 0.008 (0.000) 
& 0.223 (0.089) & 0.645 (0.039) % & 0.007 (0.000) 
& 0.312 (0.070) & 0.571 (0.050) % & 0.006 (0.001) 
& 0.462 (0.062) & 0.615 (0.039) % & 0.007 (0.000) 
& 0.373 (0.071) \\
\hline
\multirow{6}{2.5em}{Discriminant Analysis} & \multirow{2}{2.5em}{Census} 
& sex  & 0.760 (0.005) & 0.052 (0.009) % & 0.003 (0.001) 
& 0.995 (0.002) & 0.144 (0.006) % & 0.008 (0.000) 
& 0.964 (0.003) & 0.002 (0.000) % & 0.000 (0.000) 
& \textbf{1.000 (0.000)} & 0.004 (0.000) % & 0.000 (0.000) 
& \textbf{1.000 (0.000)} \\
 & & race  & 0.325 (0.012) & 0.031 (0.007) % & 0.002 (0.000) 
 & 0.991 (0.004) & 0.153 (0.015) % & 0.012 (0.001) 
 & 0.777 (0.028) & 0.003 (0.001) % & 0.000 (0.000) 
 & \textbf{1.000 (0.000)} & 0.006 (0.001) % & 0.000 (0.000) 
 & \textbf{1.000 (0.000)}\\
\cline{2-12}
 & \multirow{2}{2.5em}{Compas} 
 & sex  & 0.106 (0.002) & 0.023 (0.013) % & 0.000 (0.000) 
 & 0.937 (0.090) & 0.022 (0.002) % & 0.000 (0.000) 
 & 0.955 (0.008) & 0.001 (0.001) % & 0.000 (0.000) 
 & \textbf{1.000 (0.000)} & 0.002 (0.000) % & 0.000 (0.000) 
 & 0.999 (0.000) \\
 & &  race  & 0.139 (0.004) & 0.024 (0.006) % & 0.000 (0.000) 
 & 0.969 (0.014) & 0.023 (0.002) % & 0.000 (0.000) 
 & 0.972 (0.006) & 0.001 (0.001) % & 0.000 (0.000) 
 & \textbf{1.000 (0.000)} & 0.002 (0.000) % & 0.000 (0.000) 
 & \textbf{1.000 (0.000)} \\
\cline{2-12}
& Credit & sex  & 0.944 (0.005) & 0.097 (0.009) % & 0.006 (0.001) 
& 0.989 (0.002) & 0.288 (0.007) % & 0.018 (0.000) 
& 0.907 (0.005) & 0.006 (0.003) % & 0.000 (0.000) 
& \textbf{1.000 (0.000)} & 0.005 (0.003) % & 0.000 (0.000) 
& \textbf{1.000 (0.000)} \\
\cline{2-12}
& Bank & age  & 0.369 (0.016) & 0.186 (0.026) % & 0.005 (0.001) 
& 0.742 (0.062) & 0.260 (0.019) % & 0.008 (0.001) 
& 0.499 (0.066) & 0.023 (0.005) % & 0.001 (0.000) 
& \textbf{0.996 (0.002)} & 0.027 (0.004) % & 0.001 (0.000) 
& \textbf{0.994 (0.001)} \\
 \hline
 
\end{tabu}
}
\label{table:RQ-1}
\end{table*}
\end{flushleft}

\subsection*{RQ1: the relationship between hyperparameters and fairness (fixed dataset)}

To study the RQ1, we use 30 different benchmarks: ML hyperparameters of $5$ training algorithms, trained over
$6$ different fairness-sensitive tasks (e.g., Decision Tree over Census dataset with Sex; Decision Tree over 
Census with Race, etc.).

Table~\ref{table:RQ-1} presents the performance of various ML prediction algorithms (DNN, SVR, TR, and XGB). These algorithms are assessed based on their ability to learn the relationship between hyperparameters and fairness across the 30 scenarios. For each ML prediction algorithm, we evaluate its performance using three key metrics: Root Mean Square Error (RMSE), Relative RMSE, and the coefficient of determination ($R^2$).
Additionally, we compare these results with baseline performance, specifically in terms of Relative RMSE. It's important to note that an ML prediction algorithm is not deemed competitive if its Relative RMSE exceeds that of the baseline. Furthermore, in our assessment criteria, any $R^2$ values below 0.5 are not considered competitive.

In Table~\ref{table:RQ-1},we present the results of our analysis, which include both the average values and standard deviations from 10 repeated experiments for each of the 30 benchmarks. In our experiments, we use unique splits of training and testing fairness traces with
a ratio of 80\% training and 20\% testing ratio. The results are reported, with the standard deviations enclosed in parentheses alongside the averages for each benchmark and ML prediction algorithm. We compare the results based on these averages and the range of two standard deviations; and deem a method
outperforms another if the average is not within two standard deviations.

\begin{flushleft}
\setlength{\tabcolsep}{16.0pt}
\begin{table}[ht]
\caption{(RQ1) The performance of Tree Regressors when the EOD fairness, instead of AOD, is used. The red fonts
show benchmarks when $R^2$ is degraded whereas the bold fonts show ones that $R^2$ is increased. }
\centering
\resizebox{0.45\textwidth}{!}{%
\begin{tabu}{|l|l|l|l|l|}

\hline
\multirow{2}{4.7em}{\textbf{Algorithm}} & \multirow{2}{3.0 em}{\textbf{Dataset}} & \multirow{2}{1.5 em}{\textbf{Prot.}}  & \multirow{2}{2.0 em}{\textbf{Base.}}  &\multicolumn{1}{ l |}{\textsc{Tree Regressor}} \\ \cline{5-5}

 & & & &  $R^2$ \\
\hline
\multirow{6}{2.5em}{Decision Tree} & \multirow{2}{2.5em}{Census}
& sex  & 0.518 (0.018)  & 0.882 (0.015) \\
 & & race  & 0.691 (0.034)  & 0.868 (0.038) \\
\cline{2-5}
 & \multirow{2}{2.5em}{Compas} 
 & sex  & 11.910 (1.950)  & \textcolor{red}{0.674 (0.100)} \\
 & &  race  & 13.114 (2.590) & \textcolor{red}{0.560 (0.109)} \\
\cline{2-5}
& Credit & sex  & 2.309 (0.042)  & 0.897 (0.008) \\
\cline{2-5}
& Bank & age  & 4.083 (0.696) & 0.689 (0.074) \\
\hline
\multirow{6}{2.5em}{Logistic Regression} & \multirow{2}{2.5em}{Census} 
& sex  & 0.480 (0.013)   & 0.983 (0.007) \\
 & & race  & 0.191 (0.009)   & 0.930 (0.070) \\
\cline{2-5}
 & \multirow{2}{2.5em}{Compas} 
 & sex  & 1.198 (0.009)  & 0.999 (0.000) \\
 & &  race  & 1.913 (0.038) & 0.999 (0.000) \\
\cline{2-5}
& Credit & sex  & 0.489 (0.002)  & 0.925 (0.006) \\
\cline{2-5}
& Bank & age  & 0.735 (0.013)  & 0.933 (0.014) \\
\hline
\multirow{6}{2.5em}{Random Forest} & \multirow{2}{2.5em}{Census} 
& sex  & 0.322 (0.014) & \textbf{0.955 (0.010)} \\
 & & race  & 0.211 (0.014)  & \textbf{0.711 (0.044)} \\
\cline{2-5}
 & \multirow{2}{2.5em}{Compas} 
 & sex  & 18.002 (7.909)  & 0.553 (0.339) \\
 & &  race & 19.452 (5.010)  & 0.823 (0.047) \\
\cline{2-5}
& Credit & sex  & 0.748 (0.011)  & \textcolor{red}{0.608 (0.012)} \\
\cline{2-5}
& Bank & age  & 0.761 (0.123)  & 0.692 (0.104) \\
\hline
\multirow{6}{2.5em}{SVM} & \multirow{2}{2.5em}{Census} 
& sex  & 0.744 (0.013)  & \textcolor{red}{0.869 (0.018)} \\
 & & race  & 0.705 (0.015)  & \textcolor{red}{0.602 (0.048)} \\
\cline{2-5}
 & \multirow{2}{2.5em}{Compas} 
 & sex  & 1.185 (0.018)  & \textcolor{red}{0.813 (0.011)} \\
 & &  race  & 1.708 (0.071)  & \textbf{0.539 (0.035)} \\
\cline{2-5}
& Credit & sex  & 0.460 (0.001)  & 0.998 (0.000) \\
\cline{2-5}
& Bank & age  & 0.909 (0.058)  & \textbf{0.495 (0.067)} \\
\hline
\multirow{6}{2.5em}{Discriminant Analysis} & \multirow{2}{2.5em}{Census} 
& sex  & 0.889 (0.007)  & 1.000 (0.000) \\
 & & race  & 0.338 (0.010)& 1.000 (0.000) \\
\cline{2-5}
 & \multirow{2}{2.5em}{Compas} 
 & sex  & 11.644 (1.304) & \textcolor{red}{0.980 (0.013)} \\
 & &  race  & 13.242 (1.015) & 0.992 (0.015) \\
\cline{2-5}
& Credit & sex  & 1.000 (0.008)  & 1.000 (0.000) \\
\cline{2-5}
& Bank & age  & 0.519 (0.026)  & 0.997 (0.001) \\
 \hline
\end{tabu}
}
\vspace{-1.0 em}
\label{table:RQ-1-EOD}
\end{table}
\end{flushleft}

We highlight the best $R^2$ result as well as those within 2 standard deviations of the best result.
For example, in the Random Forest case, trained over the Credit dataset with Sex as the
protected attribute, the Tree Regressor algorithm has outperformed all other methods in predicting
the AOD fairness of Random Forest hyperparameters whereas in the benchmark with the SVM 
% \srsays{should this be SVM or SVR?} \stsays{SVM is correct}
algorithm trained over Census with Sex; DNN, Tree Regressor, and XGBoost have all achieved similar performance
to predict AOD fairness of SVM hyperparameters. Overall, we highlight TR algorithms in 93.3\% (28
out of 30 cases); XGB in 80\% (24 out of 30); DNN in 6.6\% (2 out of 30); and SVR in 0.0\% (0 out of 30).
Note that in some cases the $R^2$ measure is the same on both algorithms (e.g., see Discriminant Analysis benchmark
with Compas dataset and Race); but if we look at the Relative RMSE measure we can observe that a smaller
value is better. Therefore we can differentiate the best algorithm per dataset.
To precisely compare different ML prediction algorithms, we report the many cases when we have
$R^2$ values greater than a specific threshold.

% On Table~\ref{table:RQ-1} we have highlighted the best $R^2$ result as well as the values that are within 2 standard deviations of the best result. As mentioned before, $R^2$ is our main metric throughout our experiments. If a value is within this range then the value will be considered statistically significant since is very close to the best ML prediction algorithm score. Furthermore, values that do not meet the 0.5 threshold will not be considered statistically significant.

% \stsays{we need to talk a bit more specific about R$^2$ values...how many cases you have R2 above 0.95...how many cases above 0.8 for TR and XGB...this will convince the significance of results...}

\vspace{0.25 em}
\noindent \textbf{How many $R^2$ values do we have more than 0.95?}
Both Tree Regressor and XGBoost have 12/30 (40\%) values exceeding this threshold noting that the datasets in which $R^2$ values exceed 0.95 are the same datasets for both algorithms.
For the Deep Neural Network model, we have 6/30 (20\%) cases exceeding this threshold including 
half of the previous datasets in TR and XGB. Finally, the SVR method only has 3/30 (10\%) values that have the $R^2$
greater than 0.95.

\vspace{0.25 em}
\noindent \textbf{How many $R^2$ values do we have more than 0.80?}
Tree Regressor exceeds this threshold in 22/30 (73\%) datasets, and XGBoost is very close to it with 21/30 (70\%)
datasets meeting the 0.8 threshold. For the Deep Neural Network, 11/30 (37\%), close to half of the previous datasets meet the $R^2$ threshold. Finally, the SVR method has 6/30 (20\%) values that are greater than 0.8.

\vspace{0.25 em}
\noindent \textbf{Comparing the ML prediction algorithms using a different fairness notion.}
We also perform experiments to understand the efficacy of algorithms to predict a different
notion of fairness than AOD. Specifically, we consider the equal opportunity difference
(EOD), the true positive rate differences between two protected groups,
as the target variable of our training tasks. Since Tree Regressor has achieved the highest performance;
we are only training Tree Regressors models to predict EOD fairness. 

Table~\ref{table:RQ-1-EOD} shows the results
of experiments. In 19/30 cases (63.3\%); we observe that TR shows a similar performance when
used to predict EOD, instead of AOD. We observe that in 7/30 cases (23.3\%), the 
performance of TR (based on $R^2$) is degraded: TR predicts AOD more accurately than
EOD. When we closely examine these cases, we observe that 4 cases out of 7 involve the
COMPAS dataset. We revisited the training datasets for COMPAS and noted that the
difference in true positive rates between two groups are zero for many examples, while
this is not the case for false positive rates that contribute to AOD.
We conjecture that this is the main reason behind the difficulty in learning EOD,
compared to AOD. In 4/30 cases (13.3\%), TR predicts EOD more accurately than AOD. 

\vspace{0.25 em}
\noindent \textbf{Comparing the computational complexity of ML prediction algorithms.}
While we observe that Tree Regressor slightly outperforms XGBoost in the accuracy
of predicting fairness of ML hyperparameters; we observed that XGBoost is significantly
faster, meaning that it trains ML prediction models in relatively less running time.
Figure~\ref{fig:computational-complexity} shows the computation times of training
with TR vs. XGB over three ML training algorithms (DT, LR, and SVM), noting that
the results are similar for the other two algorithms (RF and DA). In each plot (corresponds
to one ML training algorithm), we have different datasets with a protected attribute
and the size of fairness traces used to train the regression models. As the size
grows, the XGBoost algorithm significantly scales better.

\begin{flushleft}
\setlength{\tabcolsep}{25.0pt}
\begin{table*}[ht]
\caption{Different versions of income census datasets that are released in 2014, 2015, and 2018.}
\centering
\resizebox{0.75\textwidth}{!}{%
\begin{tabu}{|l|c|c|c|l|l|l|l|}

\hline
\multirow{2}{4.7em}{\textbf{Algorithm}} & \multirow{2}{3.0em}{\textbf{Dataset}} & \multirow{2}{*}{\textbf{Prot.}}  & \multirow{2}{*}{\textbf{Baseline}} & \multicolumn{1}{| l |}{\textsc{DNN}} & \multicolumn{1}{| l |}{\textsc{SVM}}  &\multicolumn{1}{| l |} {\textsc{TR}} & \multicolumn{1}{| l |}{XGB}\\\cline{5-8}

 & & & & $R^2$  & $R^2$  & $R^2$ & $R^2$ \\
\hline
\multirow{4}{2.5em}{Decision Tree} & 2014 &   \multirow{3}{2.5em}{Sex}
& 0.672 (0.030)  & 0.372 (0.043)

&  0.287 (0.033) & \textbf{0.845 (0.043)} &\textbf{0.835 (0.045)}
\\
 & 2015&  
& 0.651 (0.022)  & 0.649 (0.041)
 
&  0.079 (0.020)  & \textbf{0.933 (0.015)} & \textbf{0.923 (0.019)}
 \\

 & 2018&  
& 0.979 (0.049)  & 0.657 (0.033) 
 
 & 0.279 (0.033) & \textbf{0.932 (0.015)}  & \textbf{0.932 (0.015)}
 \\

\cline{2-8}
 &   2014 &\multirow{3}{2.5em}{Race}
& 0.374 (0.007) & 0.386 (0.081)
 
&  0.108 (0.043) &  \textbf{0.940 (0.015)} &\textbf{0.936 (0.011)} \\
 
 &   2015& & 0.481 (0.008) &  0.664 (0.038)
 
&  0.611 (0.021) & \textbf{0.940 (0.012)} & \textbf{0.937 (0.018)} \\

 &   2018& & 0.365 (0.015) &0.454 (0.084)
 
 &0.166 (0.026) & \textbf{0.936 (0.012)} & 0.907 (0.017)  \\
\hline
\multirow{4}{2.5em}{Logistic Regression} &   2014& \multirow{3}{2.5em}{Sex} 
& 0.706 (0.015) &  0.946 (0.032) 

 & 0.930 (0.008) & \textbf{0.988 (0.006)} & \textbf{0.985 (0.007)}
\\
 &  2015&  & 0.616 (0.008) &  0.946 (0.017)
 
& 0.933 (0.014) & \textbf{0.992 (0.002)} &  \textbf{0.992 (0.002)}  \\

 &  2018& & 0.813 (0.036) & 0.910 (0.038) 
 
& 0.801 (0.038) &  \textbf{0.989 (0.008)}  & \textbf{0.990 (0.007)} \\

\cline{2-8}
 & 2014 &\multirow{3}{2.5em}{Race}  & 0.144 (0.004)  % & -0.051 (0.585)
& NV 
&0.371 (0.062) & \textbf{0.830 (0.016)} & \textbf{0.807 (0.027)}
 \\
 &   2015 & & 0.136 (0.008) & 0.382 (0.221) 
 
& 0.688 (0.070) &  \textbf{0.890 (0.031)}  & \textbf{0.898 (0.026)} \\

&   2018 && 0.161 (0.005) & 0.265 (0.171) 
 
& 0.445 (0.048) & \textbf{0.719 (0.044)}  & \textbf{0.665 (0.055)} \\
\hline
\multirow{4}{2.5em}{Random Forest} & 2014 & \multirow{3}{2.5em}{Sex} & 0.098 (0.008)  % & -0.659 (0.601) 
& NV
&  0.435 (0.100) & \textbf{0.865 (0.028)} & \textbf{0.862 (0.038)}
\\
 &  2015 && 0.197 (0.006) & 0.105 (0.425)
 
&  0.708 (0.046) & \textbf{0.967 (0.020)} & \textbf{0.966 (0.014)}
 \\
&  2018 && 0.188 (0.006) &  0.102 (0.227)
 
& 0.631 (0.059) & \textbf{0.918 (0.041)}  & \textbf{0.925 (0.040)}
 \\
\cline{2-8}
 & 2014 &\multirow{3}{2.5em}{Race}  & 0.170 (0.011) %& -0.058 (0.330) 
 & NV
& 0.688 (0.040) & \textbf{0.951 (0.012)} & \textbf{0.940 (0.025)}
 \\
 &  2015& & 0.238 (0.009)  & 0.449 (0.133)
 
&  0.588 (0.087)  & \textbf{0.968 (0.004)} & \textbf{0.966 (0.009)}
 \\

&  2018 && 0.151 (0.016) %& -0.885 (0.693)
 & NV
&  0.064 (0.367) & \textbf{0.873 (0.082)}  & \textbf{0.863 (0.109)}
 \\

\hline
\multirow{4}{2.5em}{SVM} & 2014 & \multirow{3}{2.5em}{Sex} & 0.632 (0.019) &  0.813 (0.027)

& 0.770 (0.031) & \textbf{0.962 (0.008)}  & \textbf{0.954 (0.009)}
\\
 &  2015& & 0.546 (0.014)  & 0.807 (0.029)
 
& 0.792 (0.031)& \textbf{0.956 (0.010)}  & \textbf{0.945 (0.016)} \\

 &  2018& & 0.937 (0.026)  & 0.654 (0.024)
 
&  0.570 (0.027)  & \textbf{0.813 (0.020)} & \textbf{0.800 (0.037)} \\

\cline{2-8}
 & 2014 
 & \multirow{3}{2.5em}{Race} & 0.206 (0.003)  & 0.537 (0.048)
 
& 0.554 (0.039)  & \textbf{0.750 (0.032)}  & \textbf{0.707 (0.041)}
 \\
 &  2015& & 0.257 (0.007) & 0.714 (0.035)
 
& 0.700 (0.054) & \textbf{0.870 (0.027)} & \textbf{0.853 (0.041)}
 \\

& 2018& & 0.691 (0.017) & 0.699 (0.042)
 
&  0.694 (0.052) & \textbf{0.841 (0.026)}  & \textbf{0.830 (0.029)}
 \\

\hline
\multirow{4}{2.5em}{Discriminant Analysis} &  2014 &\multirow{3}{2.5em}{Sex}  
& 0.730 (0.005) & 0.996 (0.001)
&  0.972 (0.004)  & \textbf{1.000 (0.001)}  & \textbf{1.000 (0.001)}
\\
 & 2015 &
& 0.741 (0.007)  & 0.995 (0.002)
&  0.984 (0.002)  & \textbf{1.000 (0.000)}  & \textbf{1.000 (0.000)} \\

 & 2018 &
& 0.706 (0.006)  & 0.995 (0.003)

& 0.968 (0.009)  & \textbf{0.999 (0.001)}  & \textbf{1.000 (0.001)} \\

\cline{2-8}
 &  2014
 & \multirow{3}{2.5em}{Race} & 0.102 (0.003)  & 0.865 (0.024) 
 
& 0.879 (0.007)  & \textbf{0.997 (0.002)} & \textbf{0.996 (0.001)} \\

&   2015  &
& 0.131 (0.002) & 0.919 (0.012) 
&  0.913 (0.009)  & \textbf{0.999 (0.000)} & \textbf{0.999 (0.000)} \\

&   2018&
& 0.112 (0.005)  & 0.868 (0.041) 

&  0.889 (0.012)  & \textbf{0.999 (0.000)} & \textbf{0.999 (0.000)} \\

 \hline
 
\end{tabu}
}
\label{table:RQ-2-Single-Year}
\end{table*}
\end{flushleft}

% \vspace{0.25 em}
\noindent \textbf{Performance of fairness models across multiple releases of the same dataset.} 
We aim to evaluate the performance of our fairness models across various releases of the same dataset. Specifically, we focus on the Census dataset, collecting its 2014, 2015, and 2018 versions. Each release is treated as an individual dataset, and we apply the experimental setup from RQ1 to these experiments.

The results of this analysis, detailed in Table~\ref{table:RQ-2-Single-Year} follow the same structure as in Table~\ref{table:RQ-1}. When assessing the efficacy of predicting AOD from hyperparameters, the Tree Regressor exhibited better performance, achieving the best performance in 100\% cases (30 out of 30 cases). XGB also demonstrated similar performance with  97\% in 29 out of 30 cases. Conversely, both DNN and SVR failed to achieve notable performance in any instance. The results suggest that DNN, SVR, TR, and XGB could achieve a $R^2$ score over 0.95 in 10\% (3 out of 30), 10\% (3 out of 30), 47\% (14 out of 30 cases), and 40\% (12 out of 30) respectively while considering cases with $R^2$ 0.8 and higher, these percentages are 33\% (11 out of 30), 30\% (9 out of 30), 93\% (28 out of 30 cases), and 93\% (28 out of 30).

Comparing the performance on the protected attributes Sex and Race, a clear preference for the Sex attribute emerges in our analysis. Specifically, 60\% (9 out of 15) of the datasets showed better performance with Sex as the protected attribute. Breaking it down by algorithm: all three Decision Tree datasets (2014, 2015, and 2018) performed better with Race, while Logistic Regression and Discriminant Analysis datasets uniformly favored Sex (3/3 in both cases). In the Random Forest datasets, one out of three showed better performance with Sex, and in the SVM datasets, two out of three favored Sex. This trend indicates a higher value for the Sex attribute in the majority of the cases.

% We can easily compare the performance on the two protected attributes Sex and Race. From the Decision Tree datasets all 3 of the datasets (2014, 2015, and 2018) have better performance in the Race protected attribute, 0/3 for Sex. From the Logistic Regression datasets 3/3 datasets performance is better in protected attribute Sex. From the Random Forest datasets 1/3 datasets perform better in protected attribute Sex. From the SVM datasets 2/3 datasets perform better in protected attribute Sex. From the Discriminant Analysis datasets 3/3 datasets perform better in protected attribute Sex. This totals to 60\% 9/15 datasets with better performance with the protected attribute Sex.

% Comparing performance across the two protected attributes, Sex and Race, we can see a notable trend favoring the Sex protected attribute. This analysis indicated that 60\% (9 out of 15) of the Census datasets exhibit a higher $R^2$ value on the Sex attribute than in the Race attribute. For example, all three datasets (2014, 2015, and 2018) produced from Logistic Regression and Discriminant Analysis have superior performance with the Sex datasets.

\begin{tcolorbox}[boxrule=1pt,left=1pt,right=1pt,top=1pt,bottom=1pt]
\textbf{Answer RQ1:}
Our experiences show that Tree Regressor and XGBoost achieve
the highest accuracy in predicting AOD fairness of HPs, with Tree Regressor slightly outperforming XGBoost.
However, XGBoost scales better to larger datasets, and learns the fairness functions faster. 
When a different fairness notion is used, we observe that Tree Regressor achieved similar or better performance.
We also find that the results are consistent in modeling the fairness of HPs across multiple releases of the same dataset where TR and XGB accurately model fairness in 100\% and 97\% cases. 
% \srsays{why do we comment out the last part?}
% \srsays{do we want any metrics in the answer for RQ1?}
% \stsays{addressed!}
\end{tcolorbox}
\vspace{-1.0 em}
% }

% {\footnotesize
\begin{flushleft}
\setlength{\tabcolsep}{25.0pt}
\begin{table*}[ht]
\caption{One-year data shift: training with income census 2014 and testing with 2015 one.}
\centering
\resizebox{0.75\textwidth}{!}{%
\begin{tabu}{|l|l|l|l|l|l|l|l|}

\hline
\multirow{2}{4.9em}{\textbf{Algorithm}} & \multirow{2}{3.0em}{\textbf{Dataset}} & \multirow{2}{1.5em}{\textbf{Prot.}}  & \multirow{2}{2.0em}{\textbf{Baseline}} & \multicolumn{1}{| l |}{\textsc{DNN}} & \multicolumn{1}{| l |}{\textsc{SVM}}  &\multicolumn{1}{| l |} {\textsc{TR}} & \multicolumn{1}{| l |}{XGB}\\\cline{5-8}

 & & & & $R^2$ &$R^2$ & $R^2$ &  $R^2$ \\
\hline
\multirow{4}{2.5em}{Decision Tree}  & Census 2014 $\to$ Census 2015 & \multirow{2}{2.5em}{Sex}
& 1.306 (0.036)  %& -0.238 (0.097)
& NV
 % & -0.325 (0.017) 
& NV
 % & -0.364 (0.042) 
& NV
 % & -0.454 (0.048)
& NV
\\
 & Census 2015 $\to$ Census 2014 & 
& 0.565 (0.006)  % & -60.586 (29.591)
& NV
 % & -0.584 (0.116) 
& NV
 % & -3.487 (0.841) 
& NV
 % & -3.176 (0.754)
& NV
 \\
\cline{2-8}
  & Census 2014 $\to$ Census 2015 & \multirow{2}{2.5em}{Race} 
& 0.523 (0.008)  % & -0.207 (0.094)
& NV
 
 % & -0.280 (0.034) 
& NV
 % & -0.192 (0.034) 
& NV
 % & -0.227 (0.044)
& NV
\\
 & Census 2015 $\to$ Census 2014 &  
& 0.355 (0.005)  % & -1.860 (0.340) 
 & NV
 % & -1.227 (0.068) 
& NV
 % & -1.664 (0.142) 
& NV
 % & -1.684 (0.137) 
& NV
\\
\hline
\multirow{4}{2.5em}{Logistic Regression} & Census 2014 $\to$ Census 2015 & \multirow{2}{2.5em}{Sex} 

& 0.690 (0.012) &  0.933 (0.021)

&  0.861 (0.012)  & \textbf{0.979 (0.002)} & \textbf{0.978 (0.002)}
\\
 & Census 2015 $\to$ Census 2014 & 
& 0.645 (0.007)  & 0.912 (0.024)
 
& 0.802 (0.016) & \textbf{0.971 (0.010)} & \textbf{0.968 (0.009)} \\
\cline{2-8}
  & Census 2014 $\to$ Census 2015 & \multirow{2}{2.5em}{Race} 
& 0.138 (0.008) &  0.082 (0.323)
 
 % & -0.068 (0.137) 
& NV
 & \textbf{0.617 (0.037)} & 0.523 (0.062)
 \\
 &  Census 2015 $\to$ Census 2014 & 
& 0.143 (0.004) % & -0.045 (0.392) 
& NV
 
& 0.235 (0.055)  & \textbf{0.519 (0.041)}  & \textbf{0.510 (0.045)} \\
\hline
\multirow{4}{2.5em}{Random Forest} & Census 2014 $\to$ Census 2015 & \multirow{2}{2.5em}{Sex} 
& 0.335 (0.007) % & -1.365 (0.639) 
& NV % & -1.043 (0.189) 
& NV
 % & -1.062 (0.167) 
& NV
 % & -1.040 (0.174)
& NV
\\
 & Census 2015 $\to$ Census 2014 &  
& 0.196 (0.008)  % & -6.544 (1.853)
& NV
 % & -5.784 (1.211) 
& NV
 % & -9.123 (1.790) 
& NV
 % & -8.559 (1.538)
& NV
 \\
\cline{2-8}
 & Census 2014 $\to$ Census 2015 & \multirow{2}{2.5em}{Race} 
& 0.237 (0.010)  % & -0.742 (0.430) 
& NV 
 % & -0.248 (0.111) 
 & NV
 & 0.385 (0.066) & 0.312 (0.076) 
 \\
 & Census 2015 $\to$ Census 2014 & 
 
& 0.213 (0.015)  % & -1.026 (0.688) 
& NV
 % & -1.244 (0.428) 
& NV
 % & -0.277 (0.269) 
& NV
 % & -0.212 (0.260) 
& NV
 \\
\hline
\multirow{4}{2.5em}{SVM} & Census 2014 $\to$ Census 2015 & \multirow{2}{2.5em}{Sex} 
& 0.639 (0.018)  & \textbf{0.590 (0.092)}

 & 0.422 (0.053) & \textbf{0.538 (0.067)} & \textbf{0.526 (0.071)}
\\
 & Census 2015 $\to$ Census 2014 & 
& 0.568 (0.017)  & 0.497 (0.068)
 
 & 0.332 (0.086) & \textbf{0.679 (0.036)}  & \textbf{0.636 (0.051)}  \\
\cline{2-8}
 &  Census 2014 $\to$ Census 2015 & \multirow{2}{2.5em}{Race} 
& 0.260 (0.007) & 0.448 (0.044) 

 & 0.278 (0.086)  & 0.414 (0.052)  & 0.346 (0.060) 
 \\
  &  Census 2015 $\to$ Census 2014 &  
& 0.204 (0.003)  % & -0.045 (0.071)
& NV
  % & -0.096 (0.041) 
& NV

& NV

& NV
 \\
\hline
\multirow{4}{2.5em}{Discriminant Analysis} & Census 2014 $\to$ Census 2015 & \multirow{2}{2.5em}{Sex} 
& 0.714 (0.006) & 0.983 (0.004)

 & 0.971 (0.003)  & \textbf{0.989 (0.000)} & \textbf{0.988 (0.000)}
\\
 & Census 2015 $\to$ Census 2014 & 
& 0.760 (0.007)  & 0.975 (0.005)
 
 & 0.959 (0.004) & \textbf{0.987 (0.001)}  & \textbf{0.987 (0.001)}  \\
\cline{2-8}
 &  Census 2014 $\to$ Census 2015 & \multirow{2}{2.5em}{Race} 
 
& 0.130 (0.002) & \textbf{0.513 (0.055)}

 & \textbf{0.613 (0.053)} & \textbf{0.583 (0.024)} & \textbf{0.585 (0.027)}
 \\
  &  Census 2015 $\to$ Census 2014 &  
  
& 0.105 (0.003)  & 0.262 (0.079)
 
 & 0.306 (0.044) & 0.368 (0.032) & 0.376 (0.032)
 \\
 \hline
 
\end{tabu}
}
\label{table:RQ-2-one-year-shift}
\end{table*}
\end{flushleft}
% }

\subsection*{RQ2: the relationship between hyperparameters and fairness over temporal data distribution shift}
In this study, we shift our focus to understanding the robustness of our predictive models in predicting the fairness of HPs under temporal distribution shifts. This analysis helps us to validate if our predictive models, which estimate the fairness of ML systems based on specific hyperparameter configurations, can maintain their accuracy over time as data characteristics change and ML system evolves. It is essential to assess whether the fairness of these algorithms remains consistent over time, and identify if there are systematic biases introduced due to changes in the underlying data distributions that degrade the performance of our predictive models. To do so, we focus on several Census dataset released in 2014, 2015, and 2018. This choice was motivated by the need to understand how the data may reflect social and demographic shifts over these years and how it can impact the predictability of fairness models based on a hyperparameter setting. 
% Hence, the results of the analysis are important for understanding the robustness and durability of our fairness models in dynamic environments and provide important context for their usefulness in long-term, real-world ML deployment. 
We utilize three releases of the Census dataset (2014, 2015, and 2018) and two protected attributes (Race and Sex) to analyze different temporal shifts.
Analyzing these scenarios aims to give insights into the robustness of fairness models facing distribution changes in the test data. To do so, we learn a fairness model on one distinct release of Census and then evaluate the performance of this model by testing it on another distinct release. 

\vspace{0.5 em}
\noindent \textbf{Evaluating the robustness of fairness models in one year shift.}
In this experiment, we consider two consequent years (2014, and 2015) of Census dataset releases to evaluate the robustness of fairness models in minimal distribution shifts. Table~\ref{table:RQ-2-one-year-shift} shows the results of this analysis. The structure of this table is similar to Table~\ref{table:RQ-1} except for the column $dataset$ that shows the scenario of temporal distribution in the dataset. For example, Census 2014 $\to$ Census 2015 in column $dataset$ shows that we first train the fairness model of different hyperparameter configurations on the 2014 release of the Census dataset then the 2015 release is used as the prediction dataset to evaluate the performance of the fairness model. The results in Table~\ref{table:RQ-2-one-year-shift} show that the Tree Regressor followed by XGBoost outperforms other algorithms. However, we observed a significant reduction in cases with $R^2$ greater than 0.95 (only 4 out of 20 cases while in 40\% (8 out of 20 cases) the prediction models failed to achieve positive $R^2$ scores when training and testing data distributions differ.

The results also show that when sex is the protected attribute, the Tree Regressor and XGBoost fairness models trained on the fairness traces of Logistic Regression and Discriminant Analysis robust against a one-year distribution shift in data achieving a minimum performance of 97\% in those cases.
\begin{flushleft}
\setlength{\tabcolsep}{25.0pt}
\begin{table*}[ht]
\caption{Three to four years data shift: training with census income 2014/15 and testing with 2018.}
\centering
\resizebox{0.75\textwidth}{!}{%
\begin{tabu}{|l|l|l|l|l|l|l|l|}

\hline
\multirow{2}{4.9em}{\textbf{Algorithm}} & \multirow{2}{3.0em}{\textbf{Dataset}} & \multirow{2}{1.5em}{\textbf{Prot.}}  & \multirow{2}{2.0em}{\textbf{Baseline}} & \multicolumn{1}{| l |}{\textsc{DNN}} & \multicolumn{1}{| l |}{\textsc{SVM}}  &\multicolumn{1}{| l |} {\textsc{TR}} & \multicolumn{1}{| l |}{XGB}\\\cline{5-8}

 & & & & $R^2$  & $R^2$ & $R^2$ &   $R^2$ \\
\hline
\multirow{4}{2.5em}{Decision Tree}  & Census 2014 $\to$ Census 2018 & \multirow{2}{2.5em}{Sex}
& 0.891 (0.032) % & -0.790 (0.379) 
& NV
& NV
& NV
& NV
\\
 & Census 2015 $\to$ Census 2018 & 
& 0.694 (0.004) % & -71.719 (42.797)
& NV
& NV
& NV
& NV
 \\
\cline{2-8}
  & Census 2014 $\to$ Census 2018 & \multirow{2}{2.5em}{Race} 
& 0.370 (0.015) 
& NV
& NV
& NV
& NV
\\
 & Census 2015 $\to$ Census 2018 &  
& 0.348 (0.014) 
& NV
 
& NV
& NV
& NV
\\
\hline
\multirow{4}{2.5em}{Logistic Regression} & Census 2014 $\to$ Census 2018 & \multirow{2}{2.5em}{Sex} 

& 0.740 (0.019)  & 0.933 (0.031)

&  0.810 (0.027) & \textbf{0.961 (0.010)}  & \textbf{0.958 (0.011)}
\\
 & Census 2015 $\to$ Census 2018 & 
& 0.691 (0.017) & 0.913 (0.027)
 
&  0.768 (0.036)  & \textbf{0.954 (0.011)} & \textbf{0.952 (0.012)}
\\
\cline{2-8}
  & Census 2014 $\to$ Census 2018 & \multirow{2}{2.5em}{Race} 
& 0.305 (0.013) % & -1.538 (0.434)
& NV
& NV
& NV
& NV
\\
 &  Census 2015 $\to$ Census 2018 & 
& 0.292 (0.012)  % & -1.999 (0.493)
& NV
& NV
& NV
& NV
\\
\hline
\multirow{4}{2.5em}{Random Forest} & Census 2014 $\to$ Census 2018 & \multirow{2}{2.5em}{Sex} 
& 0.319 (0.018)  % & -1.581 (0.431)
& NV
& NV
& NV
& NV
\\
 & Census 2015 $\to$ Census 2018 &  
& 0.188 (0.006) & 0.004 (0.181)
 
& NV
 & \textbf{0.648 (0.094)}  & \textbf{0.658 (0.061)}
 \\
\cline{2-8}
 & Census 2014 $\to$ Census 2018 & \multirow{2}{2.5em}{Race} 
& 0.211 (0.013) % & -0.798 (0.541) 
& NV
& NV
& NV
& NV
\\
 & Census 2015 $\to$ Census 2018 & 
 
& 0.303 (0.014)  % & -3.022 (2.464) 
& NV
& NV
& NV
& NV
\\
\hline
\multirow{4}{2.5em}{SVM} & Census 2014 $\to$ Census 2018 & \multirow{2}{2.5em}{Sex} 
& 0.919 (0.018)  % & -0.276 (0.085)
& NV
& NV
& NV
& NV
\\
 & Census 2015 $\to$ Census 2018 & 
& 0.817 (0.010)  % & -0.068 (0.067)
& NV
& NV
 & 0.069 (0.042) & 0.103 (0.040)
\\
\cline{2-8}
 &  Census 2014 $\to$ Census 2018 & \multirow{2}{2.5em}{Race} 
& 0.658 (0.013) & 0.011 (0.052) 

& NV
 & 0.026 (0.031)  & 0.108 (0.039)
 \\
  &  Census 2015 $\to$ Census 2018 &  
& 0.651 (0.014)  & 0.085 (0.053)
  & 0.076 (0.028) & 0.024 (0.050)  % & -0.028 (0.105)
& NV
 \\
\hline
\multirow{4}{2.5em}{Discriminant Analysis} & Census 2014 $\to$ Census 2018 & \multirow{2}{2.5em}{Sex} 
& 0.713 (0.007)  & 0.986 (0.003)

 & 0.968 (0.008)  & \textbf{0.990 (0.001)}  & \textbf{0.990 (0.001)}
\\
 & Census 2015 $\to$ Census 2018 & 
& 0.743 (0.008) & 0.970 (0.008)
 
 & 0.960 (0.009)  & \textbf{0.985 (0.002)}  & \textbf{0.985 (0.002)}
\\
\cline{2-8}
 &  Census 2014 $\to$ Census 2018 & \multirow{2}{2.5em}{Race} 
 
& 0.247 (0.005)  % & -2.289 (0.316)
& NV
& NV
& NV
& NV
\\
  &  Census 2015 $\to$ Census 2018 &  
  
& 0.265 (0.006) % & -3.556 (0.476)
& NV
& NV
& NV
& NV
 \\
 \hline
 
\end{tabu}
}
\label{table:RQ-2-Few-Years-Shift}
\end{table*}
\end{flushleft}
% }

\vspace{-1.0 em}
\noindent \textbf{Evaluating the robustness of fairness models for a longer temporal shifts.} 
To assess the robustness of our fairness models under conditions of increased temporal distribution shift, our experiment extends the time period from one year to three and four years. We utilize three versions of the Census datasets from 2014, 2015, and 2018, focusing on sex and race as the protected attributes. The goal of this experiment is to predict the fairness of the 2018 Census dataset using fairness models that were trained on earlier versions, specifically the 2014 and 2015 Census datasets. This approach allows us to evaluate how robust our models are against changes over time and maintain their accuracy in predicting fairness in different temporal contexts. Table~\ref{table:RQ-2-Few-Years-Shift} shows the results of these experiments, and the structure of the table is similar to Table~\ref{table:RQ-2-one-year-shift}. The results suggest that increasing the time interval of the distribution shift further increases the number of cases that failed to achieve positive $R^2$ scores with 60\%(12 out of 20 cases) compared to 40\% in one-year shifts. However, Tree Regressor and XGBoost fairness models showed strong robustness in Logistic Regression and Discriminant Analysis cases when sex is the protected attribute. Overall, Tree Regressor and XGBoost achieved the best performance in 4 out of 20 cases (20\%) with the $R^2$ scores higher than 0.95. When comparing the results in Table~\ref{table:RQ-2-one-year-shift} to the results of one-year shift in Table~\ref{table:RQ-2-one-year-shift} over the Tree Regressor,
Logistic Regression benchmark with Census 2014 $\to$ Census 2015 achieved $R^2$ of 0.979 while it achieved $R^2$ of 0.961 with only 0.018 loss in its performance
under a four year shift (Census 2014 $\to$ Census 2018).

\begin{tcolorbox}[boxrule=1pt,left=1pt,right=1pt,top=1pt,bottom=1pt]
\textbf{Answer RQ2:} 
Our results show that temporal distribution shifts can reduce the performance of our prediction models. We observed that not all models are robust to shifts. However, in 20\% of the cases, the HP prediction models remain robust against the temporal shift even on datasets with four years of gaps between training and test data. Specifically, we found that the HPs of Logistic Regression and Discriminant Analysis are more robust to the distribution shift when the protected attribute is sex. 
% We observe drastic changes in performance when the target dataset is of a different year release, especially for Decision Tree and Random Forest datasets. On the other hand, HP datasets from Logistic Regression and Discriminant Analysis are more robust and have higher $R^2$ score through small and big temporal shifts. 
% We also observe that the prediction is more robust for protected attribute sex than race.
% is easier to predict AOD fairness in datasets with protected attribute Sex than with Race, maintaining higher performance in 100\% of the cases predicting in larger temporal shifts.

\end{tcolorbox}

\vspace{-1.0 em}
\section{Discussion}
\label{sec:discussion}

\noindent \textit{Limitation}. The input dataset is arguably
the main source of discrimination in data-driven software,
however, it is known that the training process may amplify or suppress
the present discriminatory instances~\cite{zhang2020machine}. 
In this work, we focus on the training process and study
relationships between ML hyperparameters and fairness.
Our approach can serve as a mitigation strategy to avoid biased hyperparameter configurations
and pick promising ones for full training. However, our approach alone cannot eliminate all 
possible fairness bugs.
Our approach also requires a diverse set of hyperparameters and their fairness
characteristics using an automated fine-tuning algorithm. We rely on \textsc{Parfait-ML}~\cite{tizpaz2022fairness}
to generate such traces, but \textsc{Parfait-ML} relies on heuristics to generate a diverse set
of configurations and is not guaranteed
to always find interesting hyperparameters in a given time limit (4 hours in this work).
In addition, we only use two group fairness metrics ($AOD$ and $EOD$).
These metrics do not consider the distribution of different groups
and may deem software fair, while it might still harm a vulnerable community
or minority group. In general, coming up with a suitable fairness
definition is an open challenge.

\vspace{0.5em}
\noindent \textit{Threat to Validity}. To address the internal validity and ensure
our finding does not lead to an invalid conclusion, we follow established
guidelines~\cite{https://doi.org/10.1002/stvr.1486} and repeat every experiment $10$ times
and report both the average and standard deviations. 
We deem a result significant if it does not overlap with any other methods within
2 standard deviations that imply 95\% confidence in the results. However, this 
requirement might be conservative, instead, non-parametric methods and effect sizes can
be used to alleviate the conservativeness of our comparisons~\cite{chakraborty2021bias}.

To ensure that our results are generalizable and address external validity,
we perform our experiments on the hyperparameter space of five popular ML training
algorithm from scikit-learn library over six fairness-sensitive applications
that have been widely used in the fairness
literature. To learn the prediction model of ML hyperparameters for fairness,
we used four algorithms from TensorFlow, XGBoost, and scikit-learn library. 
However, it is an open problem whether these libraries, algorithms, and
applications are sufficiently representative and expressive to
cover challenging fairness scenarios and train complex models. 

\vspace{0.5em}
\noindent \textit{Usage Vision}. We envision software and AI engineers
using our framework in the training process of data-driven (ML) software 
development. Specifically, in the cross-validation stage, the users
can query the ML prediction models with a hyperparameter configuration
and receive its prediction of fairness. This avoids the need for training
the entire model with a biased hyperparameter, improves the efficiency of
the training process, and saves computational resources. We left further study
on the amounts of cost reductions as an interesting future work.

\vspace{-1.0 em}
\section{Conclusion}
\label{sec:conclusion}
The ML hyperparameters play an influential and challenging role in
the accuracy, robustness, and fairness of data-driven software. To understand
and optimize tuning hyperparameters for fairness, we propose a data-driven
framework based on regression methods to predict the fairness of ML hyperparameters.
We showed that this task is feasible for a fixed dataset
with a specific protected attribute; hence, it helps the developers save
computational resources during the training by avoiding biased hyperparameters.
We showed that generalizations under (temporal) distribution shift are feasible, but
such generalizations are limited to certain classes of ML algorithms, training
datasets, and protected attributes. 

There are multiple exciting future directions. One direction is to leverage
multi-objective algorithms to learn the frontiers of accuracy vs. fairness in the
ML hyperparameter space. Similarly, we can leverage such functions to learn
the prediction models for multiple fairness metrics such as EOD, AOD, and
statistical parity. 

\section*{Acknowledgements}
Tizpaz-Niari and Trivedi were partially supported by the NSF under grants CNS-2230060,
CCF-2317206, and CCF-2317207.

% \newpage

\bibliographystyle{ACM-Reference-Format}
\bibliography{references}

\end{document}